\documentclass[letter,twocolumn,showpacs,preprintnumbers,amsmath,amssymb]{revtex4}

\usepackage{url}
\usepackage{graphicx}

\usepackage{amsthm}
\usepackage{enumerate}
 \newtheorem{theorem}{Theorem}[section]
 \newtheorem{lemma}[theorem]{Lemma}
 \newtheorem{corollary}[theorem]{Corollary}
 \newtheorem{definition}[theorem]{Definition}

\newtheorem{example}[theorem]{Example}
\theoremstyle{plain}

\newcommand*{\avect}{\phi}

\def\diag{\mathrm{diag}}
\newcommand*{\textfrac}[2]{{#1}/{#2}}

\newcommand*{\bbC}{\mathbb{C}}
\newcommand*{\complex}{\bbC}

\newcommand*{\cH}{\mathcal{H}}

\newcommand*{\cK}{\mathcal{K}}

\newcommand*{\cM}{\mathcal{M}}

\newcommand*{\cU}{\mathcal{U}}

\newcommand*{\cX}{\mathcal{X}}

\newcommand*{\tr}{\mathsf{tr}}
\newcommand*{\id}{\openone}
\newcommand*{\powerstates}{\mathcal{P}}
\newcommand*{\wernerstates}{\mathcal{W}}

\newcommand*{\ket}[1]{|#1\rangle}
\newcommand*{\bra}[1]{\langle #1|}
\newcommand*{\proj}[1]{\ket{#1}\bra{#1}}

\newcommand*{\sign}{\mathsf{sign}}
\newcommand*{\Sym}[2]{\mathsf{Sym}^{#2}(#1)}

\newcommand*{\spec}{\mathsf{Spec}}

\newcommand{\ketbra}[1]{| #1 \rangle \langle #1 |}
\newcommand*{\Par}[2]{\mathrm{Par}(#1,#2)}

\newcommand*{\mytwirl}{\mathbb{T}}
\newcommand*{\tmap}{\mathsf{f}}

\newcommand*{\wstate}[2]{\rho_{#1}^{#2}}

\newcommand{\be}{\begin{equation}}
\newcommand{\ee}{\end{equation}}
\newcommand{\bea}{\begin{eqnarray}}
\newcommand{\eea}{\end{eqnarray}}
\newcommand{\bestar}{\begin{equation*}}
\newcommand{\eestar}{\end{equation*}}
\newcommand{\beastar}{\begin{eqnarray*}}
\newcommand{\eeastar}{\end{eqnarray*}}

\newcommand{\dd}[2]{\frac{\partial #1}{\partial #2}}

\def\complex{\mathbb{C}}
\def\U{\mathsf{U}}

\newcommand{\Cartan}{\mathsf{H}}

\newcommand*{\myrq}[1]{x^{#1}}
\newcommand*{\Upi}{\pi}

\begin{document}

\title{One-and-a-half quantum de Finetti theorems}

\author{Matthias \surname{Christandl}}
\email[]{matthias.christandl@qubit.org} \affiliation{Centre for
Quantum Computation, DAMTP,
             University of Cambridge,
             Cambridge CB3 0WA, UK}

\author{Robert \surname{K\"onig}}
\email[]{r.t.koenig@damtp.cam.ac.uk} \affiliation{Centre for
Quantum Computation, DAMTP,
             University of Cambridge,
             Cambridge CB3 0WA, UK}

\author{Graeme \surname{Mitchison}}
\email[]{g.j.mitchison@damtp.cam.ac.uk} \affiliation{Centre for
Quantum Computation, DAMTP,
             University of Cambridge,
             Cambridge CB3 0WA, UK}

\author{Renato \surname{Renner}}
\email[]{r.renner@damtp.cam.ac.uk} \affiliation{Centre for Quantum
Computation, DAMTP,
             University of Cambridge,
             Cambridge CB3 0WA, UK}

\begin{abstract}
  When $n-k$ systems of an $n$-partite permutation-invariant state
  are traced out, the resulting state can be approximated by a convex
  combination of tensor product states. This is the quantum de
  Finetti theorem. In this paper, we show that an upper bound on the
  trace distance of this approximation is given by $2\frac{kd^2}{n}$,
  where $d$ is the dimension of the individual system, thereby improving
  previously known bounds. Our result follows from a more general
  approximation theorem for representations of the unitary
  group. Consider a pure state that lies in the irreducible
  representation $U_{\mu +\nu} \subset U_\mu \otimes U_\nu$ of the
  unitary group $\U(d)$, for highest weights $\mu$, $\nu$ and
  $\mu+\nu$. Let $\xi_\mu$ be the state obtained by tracing out
  $U_\nu$. Then $\xi_\mu$ is close to a convex combination of the
  coherent states $U_\mu(g)\ket{v_\mu}$, where $g\in\U(d)$ and
  $\ket{v_\mu}$ is the highest weight vector in $U_\mu$.

  For the class of symmetric Werner states, which are invariant under
  both the permutation and unitary groups, we give a second de
  Finetti-style theorem (our ``half'' theorem). It arises from a
  combinatorial formula for the distance of certain special symmetric
  Werner states to states of fixed spectrum, making a connection to
  the recently defined shifted Schur functions~\cite{OkoOls96}. This
  formula also provides us with useful examples that allow us to
  conclude that finite quantum de Finetti theorems (unlike their
  classical counterparts) must depend on the dimension $d$. The last
  part of this paper analyses the structure of the set of symmetric
  Werner states and shows that the product states in this set do not
  form a polytope in general.
\end{abstract}

\pacs{03.67.-a, 02.20.Qs}

\date{\today}

\maketitle

\pagestyle{plain}

\section{Introduction} \label{sec-introduction}

There is a famous theorem about classical probability distributions,
the de Finetti theorem \cite{deFinetti37}, whose quantum analogue
has stirred up some interest recently. The original theorem states
that a symmetric probability distribution of $k$~random variables,
$P^{(k)}_{X_1\cdots X_k}$, that is {\em infinitely exchangeable},
i.e. can be extended to an $n$-partite symmetric distribution for
all $n > k$, can be written as a convex combination of identical
product distributions, i.e. for all $x_1, \ldots, x_k$ \be
\label{exact} P_{X_1\cdots X_k}(x_1, \ldots, x_k) = \int
P_X(x_1)\cdots P_X(x_k) d\mu(P_X), \ee where $\mu$ is a measure on
the set of probability distributions, $P_X$, of one variable. In the
quantum
analogue~\cite{stormer,HudMoo76,petz,cavesfuchsschack,fuchsschacksecond,
FuScSc04} a state $\rho^k$ on $\cH^{\otimes k}$ is said to be
infinitely exchangeable if it is symmetric (or
permutation-invariant), i.e. $\pi \rho^k \pi^\dagger=\rho^k$ for all
$\pi \in S_k$ and, for all $n> k$, there is a symmetric state
$\rho^n$ on $\cH^{\otimes n}$ with $\rho^k=\tr_{n-k} \rho^n$. The
theorem then states that
\begin{equation} \label{eq:deFin}
\rho^k=\int \sigma^{\otimes k}dm(\sigma)
\end{equation}
for a measure $m$ on the set of states on $\cH$.

However, the versions of this theorem that have the greatest promise
for applications relax the strong assumption of infinite
exchangeability~\cite{KoeRen05,Ren05}. For instance, one can assume
that $\rho^k$ is {\em $n$-exchangeable} for some specific $n>k$,
viz.~that $\rho^k=\tr_{n-k} \rho^n$ for some symmetric state
$\rho^n$. In that case, the exact statement in
equation~\eqref{eq:deFin} is replaced by an approximation
\be
\label{quantum-deF}
\rho^k \approx \int \sigma^{\otimes k}dm(\sigma),
\ee
as proved in~\cite{KoeRen05}, where it was shown that the error is
bounded by an expression proportional to $\frac{kd^6}{\sqrt{n-k}}$.

Our paper is structured as follows. In section~\ref{sec-deFin} we
derive an approximation theorem for states in spaces of irreducible
representations of the unitary group.  Our main
application of this theorem is an improvement of the error bound in the
approximation in ~(\ref{quantum-deF}) to $2\frac{kd}{n}$ for
Bose-symmetric states and to $2\frac{kd^2}{n}$ for arbitrary
permutation-invariant states. The last step from Bose-symmetry to
permutation-invariance is achieved by embedding permutation-invariant
states into the symmetric subspace, a technique which might be of
independent interest. We conclude this section with a discussion of
the optimality of our bounds and explain how our results can be generalised to permutation-symmetry with respect to an additional system.

In section~\ref{sec-Werner}, we prove the ``half'' theorem of our
title. This refers to a de Finetti theorem for a particular class of
states, the symmetric Werner states~\cite{Werner89}, which are
invariant under the action on the tensor product space of both the unitary and
symmetric groups. In order to prove our result we derive an exact
combinatorial expression for the distance of extremal
$n$-exchangeable Werner states to product states of fixed spectrum.
This has some mathematical interest because of the connection it makes
with shifted Schur functions~\cite{OkoOls96}.  It also provides us with
a rich supply of examples that can be used to test the tightness of
the bounds of the error in equation~\eqref{quantum-deF} and, in section
~\ref{sec:structurepwernerstates}, to explore the structure of the
set of convex combinations of tensor product states.

\section{On Coherent States and the de Finetti theorem}\label{sec-deFin}
\subsection{Approximation by coherent states}\label{sec-unitary}
In order to state our result we need to introduce some notation from
Lie group theory~\cite{CarterSegalMacDonald95}. Let $\U(d)$ be the
unitary group and fix a basis $\{\ket{i}\}_{i=0} ^{d-1}$ of $\mathbb{C}^d$ in order to
distinguish the diagonal matrices with respect to this basis as the
Cartan subgroup $\Cartan(d)$ of $\U(d)$. A weight vector with weight $\lambda=(\lambda_1, \ldots, \lambda_d)$, where each
$\lambda_i$ is an integer, is a vector  $\ket{v}$ in the representation $U$ of $\U(d)$ satisfying
$U(h)\ket{v}=\prod h_i^{\lambda_i}\ket{v}$, where $h_1, \ldots, h_d$
are the diagonal entries of $h \in \Cartan(d)$. We can equip the
set of weights with an ordering: $\lambda$ is said to be
(lexicographically) \emph{higher} than $\lambda'$ if $\lambda_i
>\lambda'_i$ for the smallest $i$ with $\lambda_i \neq \lambda'_i$.
It is a fundamental fact of representation theory that every
irreducible representations of $\U(d)$ has a unique highest weight
vector (up to scaling); the corresponding weights must be dominant,
i.e.~$\lambda_i \geq \lambda_{i+1}$. Two irreducible representations
are equivalent if and only if they have identical highest weights.
 It is therefore convenient to label irreducible
representations by their highest weights and write $U_\lambda$ for the
irreducible representation of $\U(d)$ with highest weight
$\lambda$. It will also be convenient to choose the normalisation of
the highest weight vector $\ket{v_\lambda}$ to be $\langle v_\lambda|v_\lambda\rangle=1$
in order to be able to view $\proj{v_\lambda}$ as a quantum state.

Given two irreducible representations $U_\mu$ and $U_\nu$ with
corresponding spaces $\cU_\mu$ and $\cU_\nu$ we can define the tensor
product representation $U_\mu \otimes U_\nu$ acting on $\cU_\mu
\otimes \cU_\nu$ by
$$ (U_\mu \otimes U_\nu) (g)= U_\mu(g) \otimes U_\nu(g),$$ for any $g
\in \U(d)$.  In general this representation is reducible and decomposes
as
 $$U_\mu \otimes U_\nu \cong \bigoplus_{\lambda} c_{\mu \nu}^\lambda
U_\lambda.$$ The multiplicities $c_{\mu \nu}^\lambda$ are known as
\emph{Littlewood-Richardson coefficients}. It follows from the
definition of the tensor product that $\ket{v_\mu} \otimes
\ket{v_\nu}$ is a vector of weight $\mu+\nu$, where
$(\mu+\nu)_i=\mu_i+\nu_i$. By the ordering of the weights, $\mu+\nu$
is the highest weight in $U_\mu \otimes U_\nu$ and $\ket{v_\mu}
\otimes \ket{v_\nu}$ is the only vector with this weight. We therefore
identify $\ket{v_{\mu+\nu}}$ with $\ket{v_\mu} \otimes \ket{v_\nu}$
and remark that $U_{\mu+\nu} $ appears exactly once in $U_\mu\otimes
U_\nu$.

Our first result is an approximation theorem for states in the spaces
of irreducible representations of $\U(d)$. Consider a normalised
vector $\ket{\Psi}$ in the space $\cU_{\mu+\nu}$ of the irreducible
representation $U_{\mu+\nu}$. By the above discussion we can embed
$U_{\mu+\nu}$ uniquely into the tensor product representation
$U_\mu\otimes U_\nu$. This allows us to define the reduced state of
$\ket{\Psi}$ on $\cU_\mu$ by $\xi_\mu=\tr_{\nu} \proj{\Psi}$. We shall
prove that the reduced state on $\cU_\mu$ is approximated by convex
combinations of rotated highest weight states:
\begin{definition}
For $g\in\U(d)$, let $\ket{v_\mu^g}:=U_\mu(g)\ket{v^\mu}$ be the rotated highest weight vector in $\cU_\mu$.
  Let $\powerstates_\mu(\complex^d)$ be the set of states of the form
  $\int \ket{v_\mu^g}\bra{v_\mu^g} dm(g)$, where $m$ is a probability
  measure on $\U(d)$.
\end{definition}

Here, the states $\ket{v_\mu^g}$, with $g \in \U(d)$, are {\em
coherent states} in the sense of~\cite{Perelomov86}.  For $d=2$ and
$\mu=(k, 0)\equiv (k)$, these states are the well-known $\mathsf{SU}(2)$-coherent
states.

In the following theorem, we use the trace distance, which is induced
by the trace norm $\|A\|:=\frac{1}{2}\tr |A|$ on the set of hermitian
operators.

\begin{theorem}[Approximation by coherent states] \label{thm:maindefinetti}
Let $\ket{\Psi}$ be in $\cU_{\mu +\nu}$ which we consider to be
embedded into $\cU_\mu \otimes \cU_\nu$ as described above. Then
$\xi_\mu= \tr_{\nu} \proj{\Psi}$ is $\varepsilon$-close to
$\powerstates_\mu(\complex^d)$, where $\varepsilon:= 2 ( 1- \frac{\dim U^d_\nu}{\dim
U^d_{\mu+\nu}} )$. That is, there exists a probability measure $m$ on
$\U(d)$ such that
\[
\|\xi_\mu-\int \proj{v_\mu^g}  dm(g)\|\leq  \varepsilon \ .
\]
\end{theorem}

\begin{proof}
By the definition of $\ket{v^g_\tau}$ and Schur's lemma, the
operators $E^g_\tau:=\dim U_\tau \proj{v^g_\tau}$, $g\in\U(d)$ 
together with the normalised uniform (Haar) measure $dg$ on $U(d)$
form a POVM on $\cU_\tau$, i.e.,
\begin{align}
\int E^g_\tau dg=\id_{\cU_\tau}\ .\label{eq:identitydecomp}
\end{align}
This allows us to write
\newcommand*{\pUmu}{w_g} 
\begin{align}
\xi_\mu =\int \pUmu \xi^g_\mu dg\ , \label{eq:rhomudecom}
\end{align}
where $\xi^g_\mu$ is the residual state on $\cU_\mu$ obtained when
applying $\{E^g_\nu\}$ to $\ket{\Psi}$, i.e.,
\[
\pUmu \xi_\mu^g=\tr_\nu ((\id_{\cU_\mu}\otimes E^g_\nu)\proj{\Psi})\ ,
\]
where  $\pUmu dg$ determines the probability of outcomes.

We claim that $\xi_\mu$ is close to a convex combination of the
states $\ket{v^g_\mu}$, with coefficients
 corresponding to the outcome
probabilities when measuring $\ket{\Psi}$ with $\{E^g_{\mu+\nu}\}$.
That is, we show that the probability measure $m$ on $\U(d)$ in the
statement of the theorem can be defined as $dm(g):=\tr(E^g_{\mu+\nu}\proj{\Psi})dg$. Our goal is thus to estimate
\begin{align*}
\|\xi_\mu-\int\tr(E^g_{\mu+\nu}\proj{\Psi})\proj{v^g_\mu} dg \|&=
\|S-\delta\|\ ,
\end{align*}
where, using~\eqref{eq:rhomudecom},
\begin{align*}
S&:=\int \pUmu\xi^g_\mu-\frac{\dim U_\nu}{\dim
  U_{\mu+\nu}}\tr(E^g_{\mu+\nu}\proj{\Psi})\proj{v^g_\mu} dg\ ,\\
\delta&:=(1-\frac{\dim U_\nu}{\dim U_{\mu+\nu}})\int
  \tr(E^g_{\mu+\nu}\proj{\Psi})\proj{v^g_\mu}dg\ .\\
\end{align*}
Because $\|\delta\|=\frac{1}{2}(1-\frac{\dim U_\nu}{\dim U_{\mu+\nu}})$, it
suffices to show that $\|S\|\leq \frac{3}{2}(1-\frac{\dim U_\nu}{\dim
  U_{\mu+\nu}})$.
Since $U_{\mu+\nu}\subset U_\mu\otimes U_\nu$ and $\ket{v_\mu} \otimes \ket{v_\nu}=\ket{v_{\mu+\nu}}$, we have\begin{align*} 
 \frac{\dim U_\nu}{\dim U_{\mu+\nu}}
\tr (E^g_{\mu+\nu}\proj{\Psi})&=\bra{v^g_\mu}\tr_\nu ((\id_\mu\otimes E^g_\nu)\proj{\Psi})\ket{v^g_\mu}\\
&=\pUmu\bra{v^g_\mu} \xi^g_\mu \ket{v^g_\mu}.\\
\end{align*}
So \be\label{S-equation}
S=\int \pUmu
\bigl(\xi^g_\mu-\proj{v^g_\mu}\xi^g_\mu\proj{v^g_\mu}\bigr) dg\ee
Now, for all operators $A, B$, we have
\[
A-BAB=(A-BA)+(A-AB)-(\id-B)A(\id-B)\ ,
\]
so putting $A=\xi_\mu^g$ and $B=\proj{v^g_\mu}$ in (\ref{S-equation}), we have
\[
S=\alpha+\beta-\gamma,
\]
where
\begin{align*}
\alpha&:= \int \pUmu(\xi^g_\mu - \proj{v^g_\mu} \xi^g_\mu) dg \\
\beta&:=  \int \pUmu (\xi^g_\mu -  \xi^g_\mu  \proj{v^g_\mu}) dg\\
\gamma&:= \int \pUmu(\id_{U_\mu} - \proj{v^g_\mu}) \xi^g_\mu
(\id_{U_\mu} - \proj{v^g_\mu}) dg.
\end{align*}
Combining $\pUmu\proj{v^g_\mu} \xi^g_\mu= \tr_{\nu} \big((\proj{v^g_\mu}\otimes
E^g_\nu)\proj{\Psi}\big)= \frac{\dim U_\nu}{\dim U_{\mu+\nu}}
\tr_\nu E^g_{\mu+\nu}\proj{\Psi}$ with~\eqref{eq:identitydecomp} and~\eqref{eq:rhomudecom}, we get 
\begin{align*}
\alpha&= \bigl(1-\frac{\dim U_\nu}{\dim U_{\mu+\nu}}\bigr) \tr_\nu
\proj{\Psi}\ .
\end{align*}
Similarly,
\begin{align*}
\beta&= \bigl(1-\frac{\dim U_\nu}{\dim U_{\mu+\nu}}\bigr)
\tr_\nu\proj{\Psi}\ ,
\end{align*}
and hence
\[ \|\alpha\|=\|\beta\| = \frac{1}{2} \left( 1 - \frac{\dim U_\nu}{\dim
U_{\mu+\nu}} \right)\ .
\]
Note that for a projector $P$ and a state $\xi$ on $\cH$, we have
\[
\frac{1}{2}\tr(P\xi)=\|P\xi P\|\ ,
\]
as a consequence of the cyclicity of the trace and the fact  that
the operator $P\xi P=(\sqrt{\xi} P)^\dagger (\sqrt{\xi} P)$  is
nonnegative. This identity together with the convexity of the
trace distance applied to the projectors
$\id_{\cU_\mu}-\proj{v^g_\mu}$ gives
\begin{align*}
\| \gamma \| &\leq \int \pUmu \| (\id_{\cU_\mu} - \proj{v^g_\mu})
\xi^g_\mu(\id_{\cU_\mu} -\proj{v^g_\mu}) \| dg  \\
&=\frac{1}{2} \tr \Bigl(
\int \pUmu (\xi^g_\mu - \proj{v^g_\mu} \xi^g_\mu) dg\Bigr)\\
& = \|\alpha \|\ .
\end{align*}
This concludes the proof because
\begin{align*}
\|\xi_\mu-\int  \tr(E^g_{\mu+\nu}\proj{\Psi})&\proj{v^g_\mu}dg\|\\
&\leq \|S\|+\|\delta\|\\
&\leq \|\alpha\|+\|\beta\|+\|\gamma\|+\|\delta\|\
\end{align*}
and each of the quantities in the sum on the r.h.s.~is upper bounded
by $\frac{1}{2}(1-\frac{\dim U_\nu}{\dim U_{\mu+\nu}})$.
\end{proof}
An important special case of Theorem~\ref{thm:maindefinetti} is the
case where $\mu=(k)\equiv(k,\underbrace{0, \ldots, 0}_{d-1})$ and $\mu+\nu=(n)$. In
this case, 
$U_{(k)}\cong\mathsf{Sym}^{k}(\complex^d)$
(and likewise for $U_{(n)}$) is the symmetric subspace of $(\complex^d)^{\otimes k}$.  Its importance stems from the fact that any $n$-exchangeable density operator has a symmetric
purification, and this leads to a new de Finetti theorem for general
mixed symmetric states (cf. Section~\ref{sec:symmetryandpurificat}).

\begin{corollary}\label{thm:symmdefinetti}
  Let $\ket{\Psi} \in \mathsf{Sym}^{n}(\complex^d)$ be a symmetric
  state and let $\xi^k:=\tr_{n-k}\ket{\Psi}\bra{\Psi}$, $k \leq n$, be
  the state obtained by tracing out $n-k$ systems. Then $\xi^k$ is
  $\varepsilon$-close to $\powerstates_{(k)}(\complex^d)$, where
  $\varepsilon:=2\frac{dk}{n}$. Equivalently, there exists a probability measure $m$ on pure states on 
  $\complex^d$ such that
\[
\|\xi^k-\int \proj{\varphi}^{\otimes k}  dm(\varphi)\|\leq  \varepsilon \ .
\]
\end{corollary}

\begin{proof}
Put $\mu=(k)$, $\nu=(n-k)$ in Theorem~\ref{thm:maindefinetti}. Then
$\ket{\Psi}\in\cU_{(n)}=\mathsf{Sym}^{n}(\complex^d)$ 
is a symmetric state,
the highest weight vector of $U_\mu$ is just the product
$\ket{0}^{\otimes k}$, and tracing out $U_\nu$ corresponds to tracing
out $(\complex^d)^{\otimes n-k}$. Since $U_\mu(g)\ket{v^\mu}=(g\ket{0})^{\otimes
k}$, an arbitrary state $\ket{\varphi}\in\complex^d$ can be written as
$g\ket{0}$ for some $g \in \U(d)$.

For the symmetric representation $U_{(l)}$, $\dim U_{(l)}=\binom{l+d-1}{l}$, so the error in the theorem is
$\varepsilon:= 2 ( 1-
\frac{\binom{n-k+d-1}{n-k}}{\binom{n+d-1}{n}})$, and
\begin{align*}
\frac{\binom{n-k+d-1}{n-k}}{\binom{n+d-1}{n}}&=\frac{(n-k+d-1)!}{(n-k)!(d-1)!}\frac{n!(d-1)!}{(n+d-1)!}\\
        &=\left( \frac{n-k+1}{n+1} \right) \cdots \left( \frac{n-k+d-1}{n+d-1} \right) \\
        &\geq  \left(\frac{n-k+1}{n+1}\right)^{d-1}\\
        &=\left(1-\frac{k}{n+1}\right)^{d-1}\\
        &\geq 1-\frac{(d-1)k}{n+1}\\
        & \geq 1- \frac{dk}{n}  \ .
\end{align*}
The first inequality here follows from $\frac{n+i}{n+k+i}\leq
\frac{n+j}{n+k+j}$, which holds for all $i\leq j$, and the second to
last inequality is also known as the `union bound' in probability
theory.
\end{proof}

\begin{example}
To get some feel for the more general case, where $U_{\mu+\nu}$ is
not the symmetric representation, let 
$1\leq p\leq d$ 
and consider
$\mu=(j^p)\equiv \underbrace{(j, \ldots, j)}_{p}$, $\nu=((m-j)^p)$
 and $\mu+\nu=(m^p)$. We can consider
the representation $U_{\mu+\nu}$ given by the Weyl tensorial
construction \cite{Weyl50}, with the tableau numbering running from
$1$ to $p$ down the first column, $p+1$ to $2p$ down the second, and
so on. Then the embedding $U_{\mu+\nu} \subset U_\mu \otimes U_\nu$
corresponds to the factoring of tensors in $(\bbC^d)^{\otimes n}=(\bbC^d)^{\otimes k} \otimes (\bbC^d)^{\otimes n-k}$, where $k=jp$ and $n=mp$.

The fact that the Young projector is obtained by symmetrising over
rows and antisymmetrising over columns implies that
$$U_{\mu+\nu} \subset \Sym{{\bigwedge}^p(\complex^d)}{m},$$
where $\bigwedge^p$ is the antisymmetric subspace on $p$~systems
corresponding to a column in the diagram. States in $\cU_{\mu+\nu}$ can
thus be regarded as symmetric states of $m$ systems of dimension
$q=\dim \bigwedge^p(\complex^d)$, and one can apply
Corollary~\ref{thm:symmdefinetti} to deduce that $\xi_\mu$ is close to
$\powerstates_{(j)}(\complex^q)$.  However,
Theorem~\ref{thm:maindefinetti} makes the assertion that $\xi_\mu$ is
close to $\powerstates_{\mu}(\complex^d)$. This statement is stronger in
certain cases.

For instance, when $p=2$, the highest weight vector
$\ket{v_\mu}$ is $(\frac{1}{\sqrt{2}}\ket{01-10})^{\otimes k}$ and
Theorem~\ref{thm:maindefinetti} says that $\xi_\mu$ is close to a convex combination of states $\proj{\varphi}^{\otimes k/2}$ where $\ket{\varphi}$ is of
the form $(g \otimes g) \frac{1}{\sqrt{2}}\ket{01-10}$ with $g \in
\U(d)$. Note that the single-system reduced density operator of
every such $\ket{\varphi}$ has rank $2$. By contrast,
Corollary~\ref{thm:symmdefinetti} allows the $\ket{\varphi}$'s to
lie in ${\bigwedge}^2(\complex^d)$, i.e. in the span of the basis elements
$\frac{1}{\sqrt{2}}\ket{i_1i_2-i_2i_1}$, for $1 \le i_1 < i_2 \le
d$. This includes
$\ket{\varphi}$'s whose reduced density operator has rank larger
than $2$, if $d > 3$.

\end{example}

\subsection{Symmetry and purification\label{sec:symmetryandpurificat}}

We now show how the symmetric-state version of our de Finetti
theorem, Corollary \ref{thm:symmdefinetti}, can be generalised to
prove a de Finetti theorem for arbitrary (not necessarily pure)
$n$-exchangeable states $\rho^k$ on $\cH^{\otimes k}$. We say a
(mixed) state $\xi^n$ on $\cH^{\otimes n}$ is {\em
permutation-invariant} or {\em symmetric} if $\Upi \xi^n
\Upi^\dagger = \xi^n$, for any permutation $\pi \in S_n$.

Here, the symmetric group $S_n$ acts on $\cH^{\otimes n}$ by
permuting the $n$ subsystems, i.e. every permutation $\pi\in S_n$
gives a unitary $\Upi$ on $\cH^{\otimes n}$ defined by \be \Upi
\ket{e_{i_1}}\otimes\cdots\otimes \ket{e_{i_n}}=
\ket{e_{i_{\pi^{-1}(1)}}}\otimes\cdots\otimes
\ket{e_{i_{\pi^{-1}(n)}}}\label{permutation-action} \ee for an
orthonormal basis $\{\ket{e_i}\}_{i=1} ^d$ of $\cH$. Note that, as a
unitary operator, $\pi^\dagger$ corresponds to the action of
$\pi^{-1}\in~S_n$.

\begin{lemma} \label{lem:sympurification}
  Let $\xi$ be a permutation-invariant state on $\cH^{\otimes n}$. Then
  there exists a purification of $\xi$ in $\Sym{\cK \otimes
    \cH}{n}$ with $\cK \cong \cH$.
\end{lemma}

\begin{proof}
  Let $A$ be the set of eigenvalues of $\xi$ and let
  $\cH_a$, for  $a \in A$, be the eigenspace of
  $\xi$, so $\xi \ket{\avect} = a \ket{\avect}$, for any
  $\ket{\avect} \in \cH_a$.  Because $\xi$ is invariant under
  permutations, we have $ \Upi^\dagger \xi \Upi \ket{\avect} =
  a \ket{\avect}$, for any $\ket{\avect} \in \cH_a$ and
  $\pi \in S_n$.  Applying the unitary operation $\Upi$ to both sides
  of this equality gives $\xi \Upi \ket{\avect} = a \Upi
  \ket{\avect}$; so $\Upi \ket{\avect} \in \cH_a$.  This
  proves that the eigenspaces $\cH_a$ of $\xi$ are invariant
  under permutations. Since the eigenspaces of $\sqrt{\xi}$ are
  identical to those of $\xi$, $\sqrt{\xi}$ is invariant under
  permutations, too. We now show how this symmetry carries over to the
  vector
  $$
  \ket{\Psi_\xi}:= (\id \otimes \sqrt{\xi}) \ket{\Psi},
  $$
  where $\ket{\Psi}=(\sum_i \ket{e_i}\otimes \ket{e_i})^{\otimes n}
  \in (\cK\otimes \cH)^{\otimes n}$ for an orthonormal basis
  $\{\ket{e_i}\}_{i=1}^d$ of $\cK \cong \cH$. Observe that
  $\ket{\Psi}$ is invariant under permutations, i.e. $(\Upi \otimes
  \Upi)\ket{\Psi}=\ket{\Psi}$. Using this fact and the permutation
  invariance of $\sqrt{\xi}$ we find
\begin{eqnarray*} (\Upi\otimes \Upi) (\id \otimes \sqrt{\xi}) \ket{\Psi} &=&  \left( \id
\otimes \Upi \sqrt{\xi} \Upi^\dagger \right) (\Upi \otimes
\Upi) \ket{\Psi}  \\
&=&(\id \otimes \sqrt{\xi})\ket{\Psi},
\end{eqnarray*}
so $\ket{\Psi_\xi}$ is invariant under permutations, and hence an
element of $\Sym{\cK \otimes \cH}{n}$.  Computing the partial
trace over $\cK^{\otimes n}$ gives
\[  \tr_{\cK^{\otimes n}} \left((\id \otimes \sqrt{\xi}) \proj{\Psi} (\id \otimes \sqrt{\xi})^\dagger \right) = \sqrt{\xi} \id \sqrt{\xi}^\dagger= \xi,\]
which shows that $\ket{\Psi_\xi}$ is a symmetric purification of
$\xi$.
\end{proof}

\begin{definition} \label{def:power}
Let $\powerstates^k=\powerstates^k(\cH)$ be the set of states of the form $\int
\sigma^{\otimes k}dm(\sigma)$, where $m$ is a probability measure on
the set of (mixed) states on $\cH$.
\end{definition}

\begin{theorem}[Approximation of symmetric states by product states]\label{thm:definettisymmetric}
  Let $\xi^{n}$ be a permutation-invariant density operator on
  $(\mathbb{C}^d)^{\otimes n}$ and $k\leq n$.  Then
  $\xi^{k}:=\tr_{n-k}(\xi^{n})$ is $\varepsilon$-close to
  $\powerstates^k(\mathbb{C}^d)$ for $\varepsilon :=2 \frac{ d^2
    k}{n}$.
\end{theorem}
\begin{proof}
  By Lemma~\ref{lem:sympurification}, there is a purification
  $\ket{\Psi}\in\Sym{\mathbb{C}^d\otimes\mathbb{C}^d}{n}$ of
  $\xi^{n}$, and the partial trace $\tr_{n-k}\proj{\Psi}$ is  $\varepsilon$-close to $\powerstates_{(k)}(\complex^{d^2})$ by Corollary \ref{thm:symmdefinetti}. The  claim then is a consequence of the fact that the trace-distance does
  not increase when systems are traced out.
\end{proof}

We close this section by looking at a stronger notion of symmetry than
permutation-invariance. This is {\em Bose-symmetry}, defined by the
condition that $\Upi\xi^n=\xi^n$ for every $\pi\in S_n$.  {\em
  Bose-exchangeability} is then defined in the obvious way. In the
course of their paper proving an infinite-exchangeability de Finetti
theorem, Hudson and Moody~\cite{HudMoo76} also showed that if $\xi^k$
is infinitely Bose-exchangeable, then $\xi^k$ is in
$\powerstates_{(k)}(\complex^d)$.  We now show that this results holds
(approximately) for Bose-$n$-exchangeable states.

\begin{theorem}[Approximation of Bose symmetric states by pure product states]\label{thm:bose}
  Let $\xi^n$ be a Bose-symmetric state on $(\complex^d)^{\otimes n}$,
  and let $\xi^k:=\tr_{n-k}(\xi^n)$, $k \leq n$. Then $\xi^k$ is
  $\varepsilon$-close to $\powerstates_{(k)}(\complex^d)$, for
  $\varepsilon:=2 \frac{dk}{n}$.
\end{theorem}

\begin{proof}
We can decompose $\xi^n$ as
\[  \xi^n= \sum_i a_i \proj{\psi_i}\]
where $\ket{\psi_i}$ is a set of orthonormal eigenvectors of
$\xi^n$ with strictly positive eigenvalues $a_i$. For all
$\pi \in S_n$ we have
\[ \Upi \ket{\psi_i} =\frac{1}{a_i} \Upi \xi^n \ket{\psi_i} = \frac{1}{a_i} \xi^n \ket{\psi_i} = \ket{\psi_i},\]
making use of the assumption $\Upi \xi^n =\xi^n$. This shows that all
$\ket{\psi_i}$ are elements of $\Sym{\complex^d}{n}$. By
Corollary~\ref{thm:symmdefinetti}, every $\xi^k_{\psi_i}=\tr_{n-k}
\proj{\psi_i}$ is $\epsilon$-close to a state $\sigma_{\psi_i}^k$ that
is in $\powerstates_{(k)}(\complex^d)$. This leads to
\[ \| \sum_i a_i \xi_{\psi_i}^k- \sum_i a_i \sigma_{\psi_i}^k\| \leq  \sum_i a_i \|\xi_{\psi_i}^k- \sigma_{\psi_i}^k\| \leq \epsilon, \]
and concludes the proof.
\end{proof}

\subsection{Optimality} \label{sec-optimality}
The error bound we obtain in Theorem~\ref{thm:definettisymmetric} is
of size $\frac{d^2k}{n}$, which is tighter than the $\frac{d^6
k}{\sqrt{n-k}}$ bound obtained in~\cite{KoeRen05}. Is there scope
for further improvement?  For classical probability distributions,
Diaconis and Freedman~\cite{DiaFre80} showed that, for
$n$-exchangeable distributions, the error, measured by the
trace distance, is bounded by
$\min\{\frac{dk}{n},\frac{k(k-1)}{2n}\}$, where $d$ is the alphabet
size. This implies that there is a bound, $\frac{k(k-1)}{2n}$, that
is independent of $d$. The following example shows that there cannot
be an analogous dimension-independent bound for a quantum de Finetti
theorem.

\begin{example}
\label{alternating-state}
Suppose $n=d$, and define a permutation-invariant state
on $(\complex^n)^{\otimes n}$ by
\[
\xi^n={1 \over n!} \sum_{\pi, \pi'} \sign(\pi)\sign(\pi') \pi
\proj{12\cdots n} \pi'^\dagger\ ,
\]
where $\{\ket{i}\}_{i=1} ^n$ is an orthonormal basis of $\mathbb{C}^n$.
This is just the normalised projector onto
$\bigwedge^n(\complex^n)$. Tracing out $n-2$ systems gives the
projector onto $\bigwedge^2(\complex^n)$, i.e. the state
\be \label{example-state}
\xi^2={2 \over n(n-1)} \sum_{1\le i<j \le n}\proj{ij-ji},
\ee
which has trace distance at least $1/2$ from
$\powerstates^2(\complex^n)$, as will be shown 
by Corollary~\ref{lem:ddependencelowerbound} and Example~\ref{example:wernerstated}.
\end{example}
We must therefore expect our quantum de Finetti error bound to
depend on $d$, as is indeed the case for the error term
$\frac{kd^2}{n}$ in Theorem~\ref{thm:definettisymmetric}.  By
generalising this example, we will show in
Lemma~\ref{lem:ddependencelowerbound} that the error term must be at
least  $\frac{d}{2n}(1-\frac{1}{d^2})$.

This example shows that some aspects of the de Finetti theorem cannot
be carried over from probability distributions to quantum states. The
following argument shows that probability distributions can, however,
be used to find lower bounds for the quantum case.

Given an $n$-partite probability distribution
$P_{X}=P_{X_1\cdots X_n}$ on $\cX^n$, define a state
\[
\ket{\Psi}:=\sum_{x\in\cX^n}\sqrt{P_{X}(x)}
\ket{x_1}\otimes\cdots\otimes\ket{x_n}\in\cH^{\otimes n}
\]
where $\{\ket{x}\}_{x\in\cX}$ is an orthonormal basis of
 $\cH$. Applying the von Neumann measurement $\cM$ defined by this
 basis to every system of $\xi^k:=\tr_{n-k}(\ket{\Psi}\bra{\Psi})$
 gives a measurement outcome distributed according to $\cM^{\otimes
 k}(\xi^k)=P_{X}$.  If $m$ is a normalised measure on
 the set of states on $\cH$, then measuring $\int \sigma^{\otimes
 k}dm(\sigma)$ gives a distribution of the form $\cM^{\otimes k} (\int
 \sigma^{\otimes k}dm(\sigma))=\int P_X^k d\mu(P_X)$.  Because the trace
 distance of the distributions obtained by applying the same
 measurement is a lower bound on the distance between two states, this
 implies that
\begin{align}\label{eq:infimum}
\inf_{\mu}\|P_{X_1\cdots X_k}-\int P_X^k d\mu(P_X)\|\leq \|\xi^k-\int
\sigma^{\otimes k}dm(\sigma)\|\ ,
\end{align}
where the infimum is over all normalised measures $\mu$ on the set of
probability distributions on $\cX$.

If $P_{X}$ is permutation-invariant, that is, if $P_{X}(x_1,\ldots,x_n)=P_{X}(x_{\pi^{-1}(1)},\ldots,x_{\pi^{-1}(n)})$
for all $(x_1,\ldots,x_n)\in\cX^n$ and $\pi\in S_n$, then
$\ket{\Psi}\in\Sym{\cH}{n}$.  Applying this to a distribution
$P_{X}$ studied by Diaconis and Freedman~\cite{DiaFre80}, and
using their lower bound on the quantity on the
l.h.s. of~\eqref{eq:infimum} gives the following result.
\begin{theorem} \label{diaconis-freedman}
There is a state $\ket{\Psi}\in\Sym{\mathbb{C}^2}{n}$ such that
the distance of $\xi^k=\tr_{n-k}\proj{\Psi}$ to $\powerstates^k$
is lower bounded by
\begin{align*}
&\frac{1}{\sqrt{2\pi e }}\cdot\frac{k}{n}+o(\frac{k}{n})\qquad
\textrm{if } n\rightarrow\infty\ \textrm{and}\ k=o(n)\\
&\phi(\alpha)+o(1)\qquad\qquad \textrm{ if } n\rightarrow\infty\
\textrm{and }\ \textfrac{k}{n}\rightarrow \alpha\in
]0,\textfrac{1}{2}[ ,
\end{align*}
where $\phi(\alpha):=\frac{1}{2\sqrt{2\pi}}\int
|1-(1-\alpha)^{\frac{1}{2}}e^{\alpha u^2/2}| e^{-u^2/2} du$.
\end{theorem}
For a fixed dimension and up to a multiplicative factor, the dependence on $k$ and $n$ in
Corollary~\ref{thm:symmdefinetti} and Theorem~\ref{thm:definettisymmetric} is therefore tight.

\subsection{De Finetti representations relative to an additional system}

A state $\xi^{A n}$ on $\cH_A \otimes \cH^{\otimes n}$ is called
  \emph{permutation-invariant} or \emph{symmetric relative to} $\cH_A$ if $({\id_A \otimes \pi}) \xi^{A n} ({\id_A
  \otimes \pi^\dagger}) = \xi^{A n}$, for any permutation $\pi \in
S_n$ (see~\cite{fannesetal,raggiowerner,KoeRen05}). This property is strictly stronger than symmetry
of the partial state $\xi^n := \tr_A(\xi^{A n})$, since symmetry of
$\xi^n$ does not necessarily imply symmetry of $\xi^{A n}$ relative to $\cH_A$, as
 the pure state
$\frac{1}{\sqrt{2}}(\ket{001}+\ket{110})\in\mathbb{C}^2\otimes(\mathbb{C}^2)^{\otimes
2}$ illustrates. Taking a broader view where $\xi^n$ is part of a
state on a larger Hilbert space thus gives rise to additional
structure.

As we shall see, this stronger notion of symmetry also yields stronger
de Finetti style statements. These are useful in applications, for
instance those related to separability problems~(cf.~\cite{Ioannou06}
and~\cite{doherty06com}, where an alternative extended de Finetti-type
theorem has been proposed). More precisely, symmetry of a state $\xi^{A n}$ on $\cH_A \otimes
\cH^{\otimes n}$ relative to $\cH_A$ implies that the partial state
$\xi^{A k} := \tr_{n-k}(\xi^{A n})$ is close to a convex combination
of states where the part on $\cH^{\otimes k}$ has product form and, in
addition, is independent of the part on $\cH_A$. In particular, $\xi^{A k}$ is close to being separable with respect to the
bipartition $\cH_A$ versus $\cH^{\otimes n}$. This property is formalised by
the following definition which generalises Definition~\ref{def:power}.

\newtheorem*{vardef}{Definition~\ref{def:power}$^\prime$}

\begin{vardef}
  Let $\powerstates^k(\cH_A,\cH)$ be the set of states of the form
  $\int \xi^A_\sigma \otimes \sigma^{\otimes k} dm(\sigma)$, where,
  $m$ is a probability measure on the set of (mixed) states on $\cH$
  and where $\{\xi^A_\sigma\}_{\sigma}$ is a family of states on
  $\cH_A$ parameterised by states on $\cH$.
\end{vardef}

The main results of Section~\ref{sec:symmetryandpurificat} can be
extended as follows.

\newtheorem*{varthma}{Theorem~\ref{thm:definettisymmetric}$^\prime$}

\begin{varthma}[Approximation of symmetric states by product states]
  Let $\xi^{A n}$ be a density operator on $\cH_A \otimes
  (\mathbb{C}^d)^{\otimes n}$ which is symmetric relative to $\cH_A$
  and let $k\leq n$.  Then $\xi^{A k}:=\tr_{n-k}(\xi^{A n})$ is
  $\varepsilon$-close to $\powerstates^k(\cH_A, \mathbb{C}^d)$ for
  $\varepsilon :=2 \frac{ d^2 k}{n}$.
\end{varthma}

A state $\xi^{A n}$ on $\cH_A \otimes \cH^{\otimes n}$ is called
\emph{Bose-symmetric relative to} $\cH_A$ if $({\id_A \otimes \pi})
\xi^{A n} = \xi^n$, for any $\pi \in S_n$.

\newtheorem*{varthmb}{Theorem~\ref{thm:bose}$^\prime$}

\begin{varthmb}[Approximation of Bose symmetric states by product states]
  Let $\xi^{A n}$ be a state on $\cH_A \otimes (\complex^d)^{\otimes
    n}$ which is Bose-symmetric relative to $\cH_A$, and let $\xi^{A
    k}:=\tr_{n-k}(\xi^{A n})$, $k \leq n$. Then $\xi^{A k}$ is
  $\varepsilon$-close to $\powerstates^{k}(\cH_A \otimes \complex^d)$,
  for $\varepsilon:=2 \frac{dk}{n}$.
\end{varthmb}

The proofs of these theorems are obtained by a simple modification of
the arguments used for the derivation of the corresponding statements
of Section~\ref{sec:symmetryandpurificat}. The main ingredient are
straightforward generalisations of Theorem~\ref{thm:maindefinetti} and
Lemma~\ref{lem:sympurification}.

\newtheorem*{varthmc}{Theorem~\ref{thm:maindefinetti}$^\prime$}

\begin{varthmc}[Approximation by coherent states]
  Let $\ket{\Psi}$ be in $\cH_A \otimes \cU_{\mu +\nu}$ and define
  $\xi_\mu:= \tr_{\nu} \proj{\Psi}$. Then there exists a probability
  measure $m$ on $\U(d)$ and a family $\{\tau_g\}_{g \in \U(d)}$ of
  states on $\cH_A$ such that
\[
\|\xi_\mu-\int \tau_g \otimes \proj{v_\mu^g}  dm(g)\|\leq  2 ( 1- \frac{\dim U^d_\nu}{\dim U^d_{\mu+\nu}} ) \ .
\] 
\end{varthmc}

\newtheorem*{varlm}{Lemma~\ref{lem:sympurification}$^\prime$}

\begin{varlm} 
  Let $\xi$ be a state on $\cH_A \otimes \cH^{\otimes n}$ which is
  permutation-invariant relative to $\cH_A$. Then there exists a
  purification of $\xi$ in $\cH_A \otimes \cK_A \otimes \Sym{\cK
    \otimes \cH}{n}$ with $\cH_A \cong \cK_A$ and $\cK \cong \cH$.
\end{varlm}

\section{On Werner States and the de Finetti theorem}\label{sec-Werner}

\subsection{Symmetric Werner states\label{sec:wernerandproducts}}

We now consider a more restricted class of states, the {\em Werner
  states} \cite{Werner89}. Their defining property is that they are
invariant under the action of the unitary group given by equation
(\ref{unitary-action}). Werner states are an interesting class of
states because they exhibit many types of phenomena, for example
different kinds of entanglement, but have a simple structure that
makes them easy to analyse.

One reason for narrowing our focus to these special states is that a
de Finetti theorem can be proved for them using entirely different
methods from the proof of Theorem~\ref{thm:maindefinetti}. We also
obtain a rich supply of examples that give insight into the
structure of exchangeable states and provide us with an
$O(\frac{d}{n})$ lower bound for
Theorem~\ref{thm:definettisymmetric}.

Schur-Weyl duality gives a decomposition
\be \label{SchurWeyl} (\mathbb{C}^d)^{\otimes k} \cong
\bigoplus_{\lambda\in \Par{k}{d}} U^d_\lambda \otimes V_\lambda\ ,
\ee
with respect to the action of the
symmetric group $S_k$ given by~\eqref{permutation-action} and the action
of the unitary group $\U(d)$ on $(\mathbb{C}^d)^{\otimes k}$ given by
\be \label{unitary-action} g \ket{\psi}=g^{\otimes k}\ket{\psi},\ \ee
for $g\in\U(d)$ and $\ket{\psi}\in (\mathbb{C}^d)^{\otimes k}$. Here $\Par{k}{d}$
denotes the set of Young diagrams with $k$ boxes and at most $d$ rows,
$U^d_\lambda$ is the irreducible representation of $\U(d)$ with
highest weight $\lambda$, and $V_\lambda$ is the corresponding
irreducible representation of $S_k$.

Let $\rho^k$ be a symmetric Werner state on $(\complex^d)^{\otimes
k}$. Schur's lemma tells us that $\rho^k$ must be proportional to
the identity on each irreducible component $U^d_\lambda \otimes
V_\lambda$, so
\be \label{werner-sum} \rho^k=\sum_\lambda w_\lambda \rho^k_\lambda\
, \ee
where $\rho^k_\lambda=P_\lambda/(\dim U^d_\lambda \dim V_\lambda)$,
with $P_\lambda$ the projector onto $U^d_\lambda \otimes V_\lambda$,
and $w_\lambda \geq 0$ for all $\lambda$, with $\sum w_\lambda=1$.

 Let $\mytwirl^k(\rho^k)$ denote the state obtained by ``twirling'' a state
$\rho^k$ on $(\mathbb{C}^d)^{\otimes k}$, i.e.,
\[
\mytwirl^k(\rho^k):=\int g^{\otimes k} \rho^k (g^{\otimes k})^\dagger dg\,
\]
where the Haar measure on $\U(d)$ with  normalisation $\int dg=1$ is used. A
state of the form $\mytwirl^k(\sigma^{\otimes k})$ is a symmetric
Werner state since its product structure ensures symmetry and twirling
makes it invariant under unitary action. We call such a state a
``twirled product state''. Any two states with the
same spectra are equivalent under twirling, so $\sigma \mapsto
\mytwirl^k(\sigma^{\otimes k})$ defines a map $\tmap^k:\spec^d \to \wernerstates^k$, where
$\spec^d$ is the set of possible $d$-dimensional spectra and $\wernerstates^k$ the
set of symmetric Werner states on $(\mathbb{C}^d)^{\otimes k}$. The map $\tmap^k$ can be
characterised as follows:

\begin{lemma}\label{lem:schur-coefficient} Given $r=(r_1,\ldots,r_d) \in \spec^d$,
the twirled product state $\tmap^k(r)$ on $(\mathbb{C}^d)^{\otimes k}$
satisfies
\[
\tmap^k(r)=\sum_{\lambda\in \Par{k}{d}} w_\lambda(r) \rho^k_\lambda\ ,
\]
where $w_\lambda(r)=\dim V_\lambda  s_\lambda (r)$ and $s_\lambda
(r)$ is the Schur function (cf.~equation~\eqref{eq:schurtableau}).
\end{lemma}
\begin{proof}
Since $\tmap^k(r)$ is a symmetric Werner state, equation~\eqref{werner-sum}
shows that it has the required form and
it remains to compute the coefficients $w_\lambda(r)$. Since the states
$\rho_\lambda^k$ are supported on orthogonal subspaces,
\[ w_\lambda(r)=\tr \ P_\lambda \tmap^k(r)\ , \]
where $P_\lambda$ is the projector onto the component $U_\lambda^d \otimes
V_\lambda$ of the Schur-Weyl decomposition of $(\mathbb{C}^d)^{\otimes k}$. Let $\sigma=\diag(r)$ be a state with spectrum $r$. By the linearity and cyclicity of the trace, 
\begin{align}
\tr(P\mytwirl^k(Q))=\tr(\mytwirl^k(P)Q)
\end{align}
for all operators $P$ and $Q$ on $(\mathbb{C}^d)^{\otimes k}$, hence we obtain
\begin{align*}
w_\lambda(r) &=\tr\left[P_\lambda \mytwirl^k(\sigma^{\otimes k})\right]\\
&=\tr\left[ \mytwirl^k(P_\lambda) \sigma^{\otimes k}\right]\\ 
&= \tr\left[ P_\lambda \sigma^{\otimes k}\right]\ .
\end{align*}
In the last step, we used the fact that $P_\lambda$ is invariant under the
action~\eqref{unitary-action}. Note that $P_\lambda$ projects onto the isotypic subspace of the
irreducible representation $U_\lambda^d$ in the $k$-fold tensor
product representation of $\U(d)$. On the one hand, this shows that
$\tr P_\lambda \sigma^{\otimes k}$ is the character of the
representation
\[ \tilde{\sigma} \mapsto P_\lambda \tilde{\sigma}^{\otimes k}
P_\lambda\ ,\]
evaluated at $\tilde{\sigma}=\sigma$.
On the other hand this
representation is equivalent to $\dim V_\lambda$ copies of
$U_\lambda^d$, whose character equals $s_\lambda(r)$. Hence,
$w_\lambda(r)=\dim V_\lambda s_\lambda(r)$.
\end{proof}

\subsection{A combinatorial formula}
We know from equation~\eqref{werner-sum}  that the states $\rho^n_\lambda$ with
$\lambda \in \Par{n}{d}$ are the extreme points of the set of
symmetric Werner states. A de Finetti theorem for the
$n$-exchangeable states
\begin{align} \tr_{n-k} \rho^n_\lambda\ ,  \qquad \mbox{for } \lambda \in
\Par{n}{d}\ ,\label{eq:extremalwernerstates}
\end{align}
therefore implies a de Finetti theorem for arbitrary
$n$-exchangeable Werner states by the convexity of the trace
distance. 

Note further that a de Finetti-type statement about all states of the
 form~\eqref{eq:extremalwernerstates}  applies to general $n$-exchangeable Werner
 states, that is,  to states  $\rho^k\in\wernerstates^k$ such that there is some symmetric state $\tau^n$ on
 $(\mathbb{C}^d)^{\otimes n}$ with $\rho^k=\tr_{n-k}\tau^n$. This is
 because we can assume that $\tau^n$ is a Werner state as
  $\rho^k=\tr_{n-k}\mytwirl^n(\tau^n)$ and $\mytwirl^n(\tau^n) \in\wernerstates^n$.

Our main step in the derivation of a de Finetti theorem for symmetric
Werner states is a combinatorial
formula for the distance of $\tr_{n-k}\rho^n_\lambda$ and the
symmetric Werner state
$\tmap^k(r)$.  Note that for every $r\in\spec^d$, the state
$\tmap^k(r)$ is a convex combination of $k$-fold product states with spectrum
 $r$, since
\begin{align} 
\tmap^k(r)= \int (g\ \diag(r)\ g^\dagger)^{\otimes k}  dg\ .\label{eq:twirledfkdef}
\end{align}

In order to present our formula for $\| \tr_{n-k}\rho^n_\lambda -\tmap^k(r)\|$,
we need to introduce the well-known Schur functions and also the
more recently defined shifted Schur functions.

We first recall the combinatorial description of the Schur function
$s_\mu$ by \be \label{eq:schurtableau}
s_\mu(\lambda_1,\ldots,\lambda_d)=\sum_T\prod_{\alpha\in\mu}
\lambda_{T(\alpha)}\ , \ee where the sum is over all semi-standard
tableaux $T$ of shape $\mu$ with entries between $1$ and $d$. A semi-standard (Young) tableau of shape $\mu$ is a Young frame filled with numbers weakly increasing to the right and strictly increasing downwards. The
product is over all boxes $\alpha$ of $\mu$ and $T(\alpha)$ denotes
the entry of box $\alpha$ in tableaux $T$. Note that
$s_\mu(\lambda)$ is homogeneous of degree $k$, where $k$ is the
number of boxes in $\mu$.

It is easy to see that the sum over semi-standard tableaux
in~\eqref{eq:schurtableau} can be replaced by a sum over all {\em
reverse tableaux} $T$ of shape $\mu$, where, in a reverse tableau,
the entries {\em decrease} left to right along each row (weakly) and
down each column (strictly). In the sequel, all the sums will be
over reverse tableaux.

The shifted Schur functions are given by the following combinatorial
formula~\cite[Theorem~(11.1)]{OkoOls96}:
\begin{align}
s^*_\mu(\lambda_1, \ldots, \lambda_d)=\sum_T\prod_{\alpha\in\mu}
(\lambda_{T(\alpha)}-c(\alpha))\ ,\label{eq:shiftedschurtableau}
\end{align}
where $c(\alpha)$ is independent of $T$ and is defined by
$c(\alpha)=j-i$ if $\alpha=(i,j)$ is the box in the $i$-th row and
$j$-th column of $\mu$.

\begin{theorem}[Distance to a twirled product state]\label{lem:Dexpression}\label{theorem:combinatorial}
Let $\lambda\in\Par{n}{d}$ and $r\in\spec^d$. 
Let $\tmap^k(r)$ be the twirled product state defined
in~\eqref{eq:twirledfkdef}. 
The distance between the partial trace
$\tr_{n-k} \rho^n_\lambda$ of the symmetric Werner state 
$\rho^n_\lambda$ and $\tmap^k(r)$  is given by
\begin{align}
\| \tr_{n-k}\rho^n_\lambda -\tmap^k(r)\|= \frac{1}{2} \sum_{\mu\in \Par{k}{d}}
\dim V_\mu |\frac{s^\star_\mu(\lambda)}{(n \downharpoonright k)}
-s_\mu(r) |\ ,\label{eq:mysumdelta}
\end{align}
where the falling factorial $(n\downharpoonright k)$ is defined to
be \newline $n (n-1) \cdots (n-k+1)$ if $k >0$ and $1$ if $k=0$.
\end{theorem}

In order to prove the theorem we will need a number of lemmas. Our first step is to express the coefficients in $\tr_{n-k}\rho^n_\lambda$ in
terms of Littlewood-Richardson coefficients.
\begin{lemma}\label{lem:youngoverlaplemma}
Let $\lambda\in\Par{n}{d}$ and let $P_\lambda$ be the projector onto $U_\lambda^d\otimes  V_\lambda$
embedded in $(\complex^d)^{\otimes n}$. Then
\[
\tr((P_\mu\otimes P_\nu)P_\lambda)=c^\lambda_{\mu\nu}\dim
U_\lambda^d \dim V_\mu \dim V_\nu
\]
for all  $\mu \in  \Par{k}{d}$ and $\nu \in \Par{n-k}{d}$, where
$c^{\lambda}_{\mu\nu}$ is the Littlewood-Richardson coefficient.
\end{lemma}
\begin{proof}
The Littlewood-Richardson coefficient $c_{\mu\nu}^\lambda$ is the
multiplicity of the irreducible representation $U_\lambda^d$ in the
decomposition of the tensor product representation ${U_\mu^d\otimes
U_\nu^d}$ of $\U(d)$, i.e.,
\begin{align}\label{eq:littlewoodrichunitary}
U_\mu^d\otimes U_\nu^d\cong \bigoplus_{\lambda}c_{\mu\nu}^\lambda
U_\lambda^d\ .
\end{align}
This implies that the image of $P_\mu\otimes P_\nu$ in
$(\mathbb{C}^d)^{\otimes n}$ is isomorphic to
\[
\bigoplus_\lambda \underbrace{\bigl(\bigoplus_{i=1} ^{c^\lambda_{\mu\nu}}
U_{\lambda,i}^d\otimes (V_{\mu,i}\otimes V_{\nu,i})\bigr)}\ ,
\]
as a representation of $\U(d)\times S_n$ where, for each $\lambda$, the underbraced part consists of
$c^\lambda_{\mu\nu}\dim V_\mu  \dim V_\nu$ copies of $U^d_\lambda$
and is contained in the component $U^d_\lambda\otimes V_\lambda$ of
the Schur-Weyl decomposition of $(\mathbb{C}^d)^{\otimes n}$. The
conclusion follows from this.
\end{proof}
Lemma~\ref{lem:youngoverlaplemma} allows us to compute the partial
trace of the projector $P_\lambda$.
\begin{lemma}\label{lem:partialtraceproj}
Let $\lambda\in\Par{n}{d}$ and let $P_\lambda$ be the projector onto $U^d_\lambda\otimes
V_\lambda$ embedded in $(\complex^d)^{\otimes n}$. Then
\[
\tr_{n-k} P_\lambda =\dim U^d_\lambda \sum_{\mu \nu} c_{\mu
\nu}^\lambda
 \frac{\dim V_\nu}{\dim U^d_\mu}P_{\mu}\ ,
\]
 where the sum extends over all $\mu \in
 \Par{k}{d}$ and $\nu \in \Par{n-k}{d}$.
\end{lemma}
\begin{proof}
Since $\tr_{n-k} P_\lambda$ is symmetric and invariant under the
action of $\U(d)$, it has the form (cf.~\eqref{werner-sum})
\[
\tr_{n-k} P_\lambda=\sum_\mu \alpha_\mu P_\mu\ .
\]
The claim then immediately follows from
\begin{align*}
\dim U^d_\mu\dim V_\mu \alpha_\mu&=\tr(P_\mu\tr_{n-k}P_\lambda)\\
&=\tr((P_\mu\otimes \id^{\otimes n-k})P_\lambda)\\
&=\tr((P_\mu\otimes \sum_\nu P_\nu)P_\lambda)
\end{align*}
and Lemma~\ref{lem:youngoverlaplemma}.
\end{proof}

In the special case where $n=k+1$ we obtain a statement that has
recently been derived by Audenaert~\cite[Proposition
4]{Audenaert04}. 

We now show how the expression for $\tr_{n-k}
P_\lambda$ in Lemma~\ref{lem:partialtraceproj} can be rewritten in
terms of shifted Schur functions. To do so we use the following
result expressing $\dim \lambda /\mu$, the number of standard
tableaux of shape $\lambda/\mu$, in terms of shifted Schur
functions.

\begin{theorem}[{\cite[Theorem
    8.1]{OkoOls96}}]\label{theorem:okols} Let $\lambda\in\Par{n}{d}$,
  $\mu\in\Par{k}{d}$ be such that $\mu_i\leq \lambda_i$ for all $i$.
  Then
\[ \frac{\dim \lambda / \mu}{\dim V_\lambda} =
\frac{s^\star_\mu(\lambda)}{(n \downharpoonright k)}\ .\]
\end{theorem}
Okounkov and Olshanski give a number of proofs for this theorem,
the second of which only uses elementary representation theory.

The shifted Schur functions allow us to express partial traces of Werner states
in a form analogous to Lemma~\ref{lem:schur-coefficient}.
\begin{lemma}\label{lem:wernerpartialtrace}
Let $\lambda\in\Par{n}{d}$.
The partial trace of the symmetric  Werner state $\wstate{\lambda}{n}$
 on
$(\mathbb{C}^d)^{\otimes n}$  satisfies
\[
\tr_{n-k} \wstate{\lambda}{n}=\sum_{\mu\in\Par{k}{d}}
\alpha^\lambda_{\mu} \wstate{\mu}{k}\
\]
where
\[
\alpha ^\lambda_{\mu}= \dim V_\mu \frac{s_\mu^\star(\lambda)}{(n \downharpoonright k)}\ .
\]
\end{lemma}
\begin{proof}
Lemma~\ref{lem:partialtraceproj} gives
\[
\alpha ^\lambda_{\mu}= \dim V_\mu
\sum_{\nu\in\Par{n-k}{d}} c^\lambda_{\mu\nu}\frac{\dim
V_\nu}{\dim V_\lambda}\ .
\]
Note that $c^\lambda_{\mu\nu}=0$  (by the Littlewood-Richardson rule)
 and $s_\mu^\star(\lambda)=0$ (by~\cite[Theorem~3.1]{OkoOls96}) unless
 $\mu_i\leq\lambda_i$ for all $i$. The claim therefore follows from Theorem~\ref{theorem:okols}
and the identity (see~\cite[p.~67]{fulton97})
\begin{align*}
\dim \lambda/\mu=\sum_{\nu\in\Par{n-k}{d}} c^\lambda_{\mu\nu}\dim V_\nu\
.
\end{align*}\qedhere
\end{proof}
We are now ready to give the proof of the combinatorial formula.
\begin{proof}[Proof of Theorem~\ref{lem:Dexpression}]
This is an immediate consequence of
Lemmas~\ref{lem:schur-coefficient} and~\ref{lem:wernerpartialtrace},
since
\begin{align*}
\| \tr_{n-k}\rho^n_\lambda-\tmap^k(r)\| & = \| \sum_\mu \alpha^\lambda_\mu \rho_\mu^k - \sum_\mu w_\mu(r) \rho_\mu^k \|\\
& = \frac{1}{2} \sum_{\mu} |\alpha^\lambda_\mu- w_\mu(r) |\ ,
\end{align*}
where we used the fact that the support of the $\rho_\mu^k$'s is
orthogonal.
\end{proof}

\subsection{A de Finetti theorem for Werner states} \label{sec-defin-werner}

The following de Finetti style theorem is a consequence of
Theorem~\ref{theorem:combinatorial}. We call it ``half a theorem'' as it is a quantum de Finetti theorem for a
restricted class of quantum states, the Werner states.

\begin{theorem}[Approximation by twirled
  products]\label{theorem:werner-expansion}
Let $\lambda\in\Par{n}{d}$ and define
$\bar{\lambda}:=(\frac{\lambda_1}{n},\ldots,\frac{\lambda_d}{n})\in\spec^d$.
Let $\tmap^k(\bar{\lambda})$ be defined as
in~\eqref{eq:twirledfkdef}. Then the partial trace
$\tr_{n-k}\rho^n_\lambda$ of the symmetric
Werner state $\rho^n_\lambda$ satisfies
\[
||\tr_{n-k}\rho^n_\lambda -\tmap
^k(\bar\lambda)|| \leq\frac{3}{4}\cdot
\frac{k(k-1)}{\lambda_\ell}+O(\frac{k^4}{\lambda_\ell^2})\ ,
\]
where $\lambda_\ell$ is the smallest non-zero row of $\lambda$. 
\end{theorem}
The
dimension $d$ does not appear explicitly in this bound, nor in the
order term $O(\cdot)$.
\begin{proof}
First note that we can restrict the sum to diagrams $\mu$ with no
more than $\ell$ rows, since by definition of $\ell$, $\lambda_q=0$
for $q>\ell$, and $s_\mu(\lambda_1, \ldots, \lambda_\ell, 0, \ldots,
0)=s^\star_\mu(\lambda_1, \ldots, \lambda_\ell, 0, \ldots, 0)=0$ for
$\mu_{\ell+1}>0$. Furthermore, Schur as well as shifted Schur
functions satisfy the stability condition~\cite{OkoOls96}
\begin{align*}
s_\mu(\lambda_1, \ldots, \lambda_\ell, 0, \ldots, 0)&=
s_\mu(\lambda_1, \ldots, \lambda_\ell)\\
s_\mu^\star(\lambda_1, \ldots, \lambda_\ell, 0, \ldots, 0)&=
s_\mu^\star(\lambda_1, \ldots, \lambda_\ell)\ ,
\end{align*}
so that we can safely assume that $\lambda$ has $\ell$
(non-vanishing) rows and that the tableaux are numbered from 1 to
$\ell$ only. Note that
\begin{align} \label{harpoonestimate}
\frac{1}{(n \downharpoonright k)}=n^{-k}
(1+\frac{k(k-1)}{2n}+O(\frac{k^4}{n^2})).
\end{align}
and
\begin{align*}
n^{-k}s_\mu^\star(\lambda)&=\sum_T\prod_\alpha
(\bar{\lambda}_{T(\alpha)}-\frac{c(\alpha)}{n})\\
&= \sum_T \prod_{\beta}\bar{\lambda}_{T(\beta)} \Big(1- \sum_\alpha
\frac{c(\alpha)}{\lambda_{T(\alpha)}}  \\
& \qquad \quad + \frac{1}{2}\prod_{\alpha\neq  \alpha'}
\frac{c(\alpha)c(\alpha')}{\lambda_{T(\alpha)}\lambda_{T(\alpha')}}+
\cdots \Big)\end{align*}
 where we have made use
of~\eqref{eq:shiftedschurtableau} in the first line.
Using~\eqref{eq:schurtableau}, the bound $|c(\alpha)| \leq k-1$ and
the fact that $\alpha$ enumerates $k$ boxes, we find the bounds
\begin{align*}
|n^{-k}s_\mu^\star(\lambda)- s_\mu(\bar{\lambda})| \leq
s_\mu(\bar{\lambda})\left( \frac{k
(k-1)}{\lambda_\ell}+O(\frac{k^4}{\lambda_\ell^2}) \right)
\end{align*}

Combining this with the estimate~\eqref{harpoonestimate} we obtain
\begin{align}\label{eq:smubarsum}
\frac{1}{2}\Bigl|\frac{s_\mu^\star(\lambda)}{(n \downharpoonright
  k)s_\mu(\bar{\lambda})}-1\Bigr|\leq \frac{3}{4}\frac{k(k-1)}{\lambda_\ell}  +O\Big(\frac{k^4}{\lambda_\ell^2}\Big),
\end{align}
where we have used $\lambda_\ell \leq n$. Since $\|\tr_{n-k}\rho^n_\lambda-\tmap^k(\bar{\lambda})\|$ is a convex combination with weights
$\tr(P_\mu\tmap^k(\bar{\lambda}))=\dim V_\mu s_\mu(\bar{\lambda})$ of
the terms on the l.h.s. of~\eqref{eq:smubarsum}, this concludes the proof.
\end{proof}
\begin{example}
Three special cases may be noted:
\begin{itemize}
\item Fix $\bar\lambda$ and consider $\lambda=n\bar\lambda$ for an
integer $n$. The bound then turns into
\[ O(\frac{k^2}{n}) \]
just as in the classical case. Thus when one restricts attention to a particular diagram shape
$\bar\lambda$, one obtains the same type of dimension-independent bound as Diaconis and Freedman~\cite{DiaFre80}. (This does not contradict Example \ref{alternating-state} where we focus on single diagram with $\lambda_\ell=1$. The bound of Theorem~\ref{theorem:werner-expansion} gives no information here.)
\item For $\lambda=(\sqrt{n}, \ldots, \sqrt{n})$ we have an error of order
\[ O(\frac{k^2}{\sqrt{n}})\ . \]
\item Finally, $\lambda=(n)$: In this case,
  $\tr_{n-k}\rho^n_\lambda=\tmap^k(1,0,\ldots,0)$ which means that
  $\tr_{n-k}\rho^n_\lambda$ has a product form and an application of
  Theorem~\ref{theorem:werner-expansion} is not needed.
\end{itemize}
\end{example}

Note that in Theorem~\ref{theorem:werner-expansion} we only kept the
dependence on the last nonzero row $\lambda_\ell$ of $\lambda$. For
specific applications (or for cases such as $\lambda= (\lambda_1,
\ldots, \lambda_{\ell-1}, 1)$) one may want to derive bounds
that depend on more details of $\lambda$.

By the (infinite) quantum de Finetti theorem, convex
combinations of tensor product states are the same thing as infinitely
exchangeable states. In this light, a finite de Finetti theorem 
says how close $n$-exchangeable states are to
$\infty$-exchangeable states, and one can generalise the notion of a
de Finetti theorem, and ask
\begin{quote}
How well can $n$-exchangeable
states be approximated by $m$-exchangeable states, where $m\geq n$?
\end{quote}

In the realm of symmetric Werner states, this amounts to bounding the distance
$\|\tr_{n-k}\rho^n_{n\bar{\lambda}}-\tr_{m-k}\rho^m_{m\bar{\lambda}}\|$,
which is
\begin{align*}
\frac{1}{2}\sum_{\mu\in \Par{k}{d}} \dim V_\mu |\frac{s^\star_\mu(n\bar{\lambda})}{(n
\downharpoonright k)} -\frac{s^\star_\mu(m\bar{\lambda})}{( m
\downharpoonright k)} |\ .
\end{align*}
A straightforward calculation very similar to the proof of
Theorem~\ref{theorem:werner-expansion} leads to an interpolation between the trivial case where $m$ equals $n$ and the case where $m\rightarrow\infty$ which we have considered in
Theorem~\ref{theorem:werner-expansion}.

\subsection{Necessity of $d$-dependence}
We end this section with a lower bound, which is a direct corollary
to Theorem~\ref{theorem:combinatorial}.
\begin{corollary}\label{lem:ddependencelowerbound}
Let $k<d$ and let $\lambda=(m^d)$ be the diagram consisting of
$d$~rows of length~$m$. Then the distance of $\tr_{n-k}\rho^n_\lambda$ to
$\powerstates^k$ is lower bounded by $\frac{d}{2(n-1)}(1-\frac{1}{d^2})$,
where $n=md$.
\end{corollary}
Note that this bound can be seen as a generalisation of
Example~\ref{alternating-state}, where we set $d=n$.
 It implies that any
quantum de Finetti theorem can only give an interesting
 statement if $d$ is small compared to $n$.
\begin{proof}
Note first that the functions $s_\mu(\bar{\lambda})$ and
$s_\mu^\star(\lambda)$ take a particularly simple form for the diagram
$\lambda$ under consideration. From equation~\eqref{eq:schurtableau}
\begin{align}
s_\mu(\bar{\lambda})=d^{-k}\dim U_\mu^k\ ,\label{eq:smuevaluated}
\end{align}
since $\dim U_\mu^k$ is equal to the number of semi-standard tableaux
$T$ of shape $\mu$, and from equation~\eqref{eq:shiftedschurtableau}
\begin{align}
n^{-k}s_\mu^\star(\lambda) &=d^{-k} \dim U_\mu^k  \prod_{\alpha}
\bigl(1-\frac{c(\alpha) d}{n}\bigr)\ .\label{eq:smustarcomputed}
\end{align}
Because the trace distance does not increase when tracing out systems,
and $\tr_{k-2} \tau^k\in\powerstates^2$ for every
$\tau^k\in\powerstates^k$, we can bound the distance of
$\tr_{n-k}\rho^n_\lambda$ to $\powerstates^k$ as follows:
\[
\min_{\tau^k\in\powerstates^k} \|\tr_{n-k}\rho^n_\lambda-\tau^k\|\geq
\min_{\tau^2\in\powerstates^2} \|\tr_{n-2}\rho^n_\lambda-\tau^2\|\ .
\]
Let $\mu=(1^2)$. We show below that
\begin{align}\label{eq:smubarcomputed}
\max_{r} s_\mu(r)=s_\mu(\bar{\lambda})\ ,
\end{align}
where the maximisation ranges over all spectra. With $\dim V_\mu=1$, this gives for every
$\tau^2\in\powerstates^2$
\begin{align*}
\|\tr_{n-2}\rho^n_\lambda-\tau^2\|
&\geq \tr (P_\mu(\tr_{n-2}\rho^n_\lambda-\tau^2))\\
&\geq \tr(P_\mu\tr_{n-2}\rho^n_\lambda)-\max_\sigma
\tr(P_\mu\sigma^{\otimes 2})\\
&\geq \frac{s_\mu^\star(\lambda)}{(n
  \downharpoonright 2)}-s_\mu(\bar{\lambda})\ ,
\end{align*}
by Lemma~\ref{lem:wernerpartialtrace}  and
Lemma~\ref{lem:schur-coefficient}.
Equation~\eqref{eq:smustarcomputed} implies
\begin{align}\label{eq:smustarcomputedsecond}
n^{-2}s_\mu^\star(\lambda)=d^{-2}\dim
U_\mu^2 (1+\frac{d}{n})\ .
\end{align}
We thus obtain
\begin{align}
\frac{s_\mu^\star(\lambda)}{(n \downharpoonright
  2)}- s_\mu(\bar{\lambda})&=
d^{-2}\dim U_\mu^2
\Bigl(\frac{1}{1-\frac{1}{n}}(1+\frac{d}{n})-1\Bigr)\nonumber\\
&= \dim U_\mu^2\frac{d+1}{(n-1)d^2}\label{eq:inequalitydiff}
\end{align}
by~\eqref{eq:smuevaluated} and~\eqref{eq:smustarcomputedsecond}. The
claim then immediately follows from $\dim U_\mu^2=\binom{d}{2}$.

It remains to prove~\eqref{eq:smubarcomputed}.
According to definition~\eqref{eq:schurtableau}, for $\mu=(1^2)$,
\[
s_{\mu}(r_1,\ldots,r_d)=\sum_{i_1<i_2} r_{i_1} r_{i_2}
\]
where the sum is over all indices $i_1,i_2\in\{1,\ldots,
d\}$. We claim that
\begin{align}\label{eq:smuinequalityonetwo}
s_\mu(r_1,\ldots,r_d)\leq s_\mu
\bigl(\frac{r_1+r_2}{2},\frac{r_1+r_2}{2},r_3,\ldots,r_d\bigr)\ .
\end{align}
This follows from the fact that we can write
\[
s_\mu(r)=r_1 r_2+(r_1+r_2)\sum_{i\geq 3}r_i
+\sum_{3\leq i_1<i_2}r_{i_1}r_{i_2}
\]
 and the inequality
\[
\sqrt{r_1r_2}\leq \frac{r_1+r_2}{2}
\]
relating the geometric and the arithmetic mean of $r_1$, $r_2$.
Inequality~\eqref{eq:smuinequalityonetwo} and the symmetry of $s_\mu$
imply~\eqref{eq:smubarcomputed}.

\end{proof}

\section{The structure of $\powerstates^k$ for Werner
  states\label{sec:structurepwernerstates}}
We focus now on the set
$\mytwirl^k(\powerstates^k)=\powerstates^k\cap\wernerstates^k$ of Werner
states that are convex combinations of product states.
Theorem~\ref{lem:Dexpression} approximates elements of
$\wernerstates^k$ by elements of $\powerstates^k
\cap\wernerstates^k$. We can ask whether it is possible for a
symmetric Werner state to be closer to $\powerstates^k$ than to the
set $\powerstates^k \cap\wernerstates^k$. The negative answer is
given by the following lemma.
\begin{lemma}
\label{prod-distance} The closest state $\tau^k\in\powerstates^k$ to
a symmetric Werner state $\rho^k\in\wernerstates^k$ is itself a Werner
state, i.e., an element of $\powerstates^k\cap \wernerstates^k$. 
\end{lemma}
\begin{proof}
Suppose $\tau^k\in\powerstates^k$ is the nearest (not necessarily Werner) product
state, so $\|\rho^k-\tau^k\|$ is minimal. Then, using the convexity of
the distance
\beastar
\|\rho^k-\mytwirl^k(\tau^k)\|&=&\|\mytwirl^k(\rho^k-\tau^k)\|\\
&\le& \int \| g^{\otimes k}(\rho^k-\tau^k)(g^\dagger)^{\otimes k}\|
dg\\ &=& \|\rho^k-\tau^k\|, \eeastar
so the Werner state $\mytwirl^k(\tau^k)$ is at least as close to
$\rho^k$ as $\tau^k$ (and in fact the triangle inequality is strict
unless $\tau$ is $\U(d)$-invariant). This means that the closest state
$\tau^k$ is an element of $\powerstates^k\cap
\wernerstates^k$. 
\end{proof}
A symmetric Werner state $\rho^k_\lambda$, for $\lambda\in\Par{k}{d}$, has
the following optimality property:
\begin{lemma}\label{lem:wernerstateclosest}
Let $\lambda\in\Par{k}{d}$. The state $\rho^k_\lambda\in\wernerstates^k$ is
closer to $\powerstates^k$ than any other state $\rho^k$ with support on
$U^d_\lambda \otimes V_\lambda$.
\end{lemma}

\begin{proof}
Let $\rho^k$ be a state with support on $U^d_\lambda\otimes
V_\lambda$, and let $\tau^k\in\powerstates^k$ be the state that is closest to $\rho^k$. By
Schur's Lemma, $\rho^k_\lambda=\mytwirl^k(\Sigma(\rho^k))$, where
$\Sigma(\rho^k):=\frac{1}{k!}\sum_{\pi\in S_k} \pi \rho^k
\pi^\dagger$. Thus, using the triangle inequality and the unitary
invariance of the trace norm, \beastar \| \rho^k_\lambda -
\mytwirl^k(\tau^k)\| &=&  \| \mytwirl^k(\Sigma(\rho^k)) -
\mytwirl^k(\Sigma(\tau^k))\|
\\ &\leq& \| \rho^k-\tau^k\|\ . \eeastar
\end{proof}

The set $\mytwirl^k(\powerstates^k)$ is the convex hull of all twirled
tensor products $\mytwirl^k(\sigma^{\otimes k})$, which is the convex
hull of $\tmap^k(\spec^d)$. Since $\tmap^k(\pi r)=\tmap^k(r)$, for any
permutation $\pi$ of $r_1, .., r_d$, we can restrict $\tmap^k$ to the
simplex $\Delta(d)=\spec^d/S_d$. The vertices of $\Delta(d)$ are the
points $\myrq{q}\in \spec^d$ whose first $q$ coordinates are $1/q$
and the remainder zero, for $q=1,....,d$. Thus $\tmap^k(\myrq{1})$ is
just the twirl of $\ketbra{0}^{\otimes k}$, which is the projector
onto $\Sym{\cH}{k}$, and $\tmap^k(\myrq{d})=(\openone/d)^{\otimes k}$
is the fully mixed state.

\begin{figure}
\centerline{\includegraphics[scale=0.5]{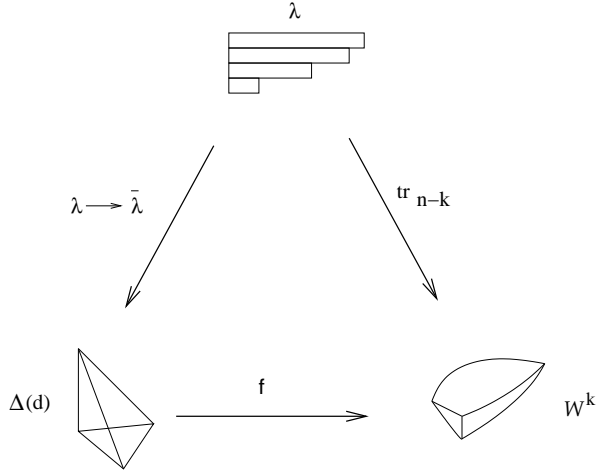}}
\caption{This shows schematically the maps underlying Theorem~\ref{theorem:werner-expansion}. From the Young diagram $\lambda$ on
$n$ systems, one can go to $\wernerstates^k$ by taking the state $\rho_\lambda^n$ and tracing out $n-k$ systems, or one can go to $\Delta(d)$ by normalising the row lengths
of $\lambda$. From the latter space, the map $\tmap^k$ takes one to
$\wernerstates^k$, and the two routes approximately end up at the same point.
\label{scheme}}
\end{figure}

The set of Werner states in $\powerstates^k$ is thus the convex hull
of $\tmap^k(\Delta(d))$ (see Figure~\ref{scheme}). What does this set
look like?

\begin{example}\label{example:wernerstated}
Let us look first at the case where $k=2$ and $d$ is arbitrary. By
Lemma \ref{lem:schur-coefficient}, the point $r$ in $\Delta(d)$ is
mapped to $s_{(2)}(r)\rho_{(2)}+s_{(1^2)}(r)\rho_{(1^2)}$, and it is
easy to check that $s_{(1^2)}(r)=\sum_{i<j}r_ir_j$ is maximised by
$r=\myrq{d}$, giving $s_{(1^2)}(r)=1/2(1-1/d)$. The states in
$\tmap^2(\Delta(d))$ are therefore those of the form
$a\rho_{(2)}+b\rho_{(1^2)}$ with $b\le 1/2(1-1/d)$. Thus $\powerstates^2\cap\wernerstates^2$ has a rather
trivial polytope structure, being a line segment.
It follows that
the state $a\rho_{(2)}+b\rho_{(1^2)}$ lies at a distance
$\max(0,b-1/2(1-1/d))$ from $\powerstates^2\cap \wernerstates^2$. By
Lemma~\ref{prod-distance}, this is also the minimum distance to
$\powerstates^2$. This result implies that the state
$\xi^2=\rho_{(1^2)}$ considered in Example~\ref{alternating-state},
equation~(\ref{example-state}), has distance at least $1/2$ from
$\powerstates^2$, showing the impossibility of a dimension-free bound
on the error of a quantum de Finetti theorem (see remarks following Corollary~\ref{lem:ddependencelowerbound}). In fact,
Lemma~\ref{lem:wernerstateclosest} implies that any symmetric state
with support on $\Lambda^2(\mathbb{C}^d)$ has distance at least
$\frac{1}{2}$ to $\powerstates^2$.
\end{example}
\begin{example}\label{example:wernerstatesvector}
Consider next the case $d=3$, $k=3$. We will henceforth  regard
the set of Werner states $\wernerstates^3$ as a subset of $\mathbb{R}^3$
by identifying a state $\rho=\sum_\lambda v_\lambda \rho_\lambda$ with
the vector $v=(v_{(3)},v_{(2,1)},v_{(1^3)})$.
If $\sigma=r_1
\ketbra{1}+r_2\ketbra{2}+r_3\ketbra{3}$, Lemma
\ref{lem:schur-coefficient} tells us that
\bea \label{tmap_r}
\tmap^3(r)&=&(s_{(3)}(r),2s_{(2,1)}(r),s_{(1^3)}(r))\\ \nonumber
&=&(\sum r_i^3+\sum_{i\ne j}r_i^2r_j +r_1r_2r_3, \\ \nonumber
&&2\sum_{i \ne j} r_i^2r_j +4r_1r_2r_3, r_1r_2r_3).
\eea
The vertices of $\Delta(d)$ are mapped to
\beastar
&&\tmap^3(x^1)=(1,0,0),\\
&&\tmap^3(x^2)=\left({1 \over 2}, {1 \over 2},0\right),\\
&&\tmap^3(x^3)=\left({10 \over 27},{16 \over 27},{1 \over 27}\right),
\eeastar
and comparison of equation (\ref{tmap_r}) and the coordinates of the vertices gives
\beastar
\tmap^3(r)&=&(\sum r_i^3- \sum r_i^2
r_j+3 r_1r_2r_3)\tmap^3(x^1)\\
&+&(4\sum r_i^2
r_j-24 r_1r_2r_3)\tmap^3(x^2)+(27 r_1r_2r_3)\tmap^3(x^3).
\eeastar
and one can show that the polynomial coefficients are positive. So
$\tmap^3(r)$ lies in the convex span  of
$\{\tmap^3(\myrq{1}), \tmap^3(\myrq{2}), \tmap^3(\myrq{3})\}$.
Note that $\tmap^3(\Delta(d))$ is a subset of the set of triseparable
Werner states studied in~\cite{eggelingwerner00}.
\end{example}

\begin{figure}
\centerline{\includegraphics[width=0.25\textwidth]{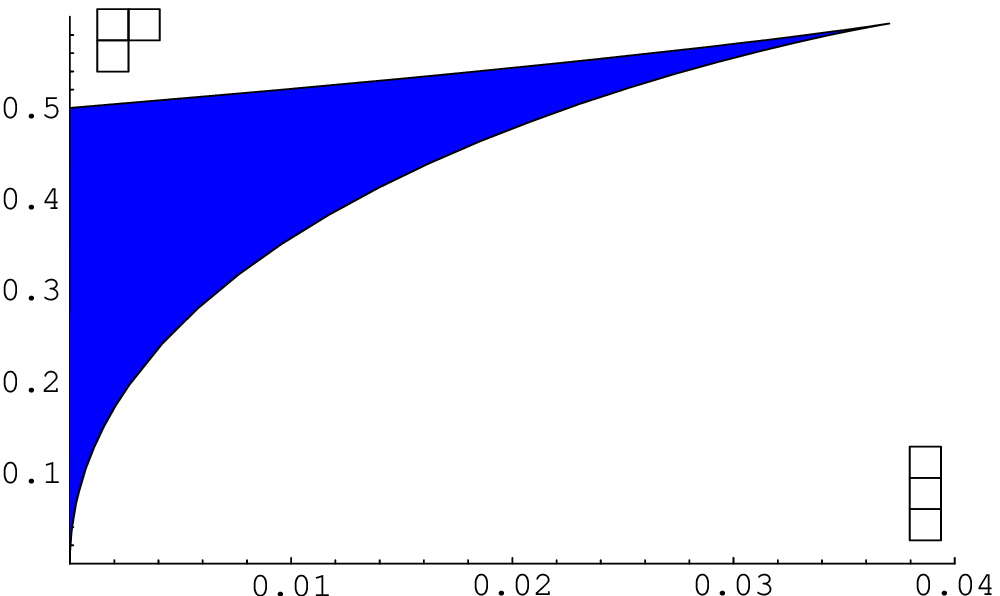}\includegraphics[width=0.25\textwidth]{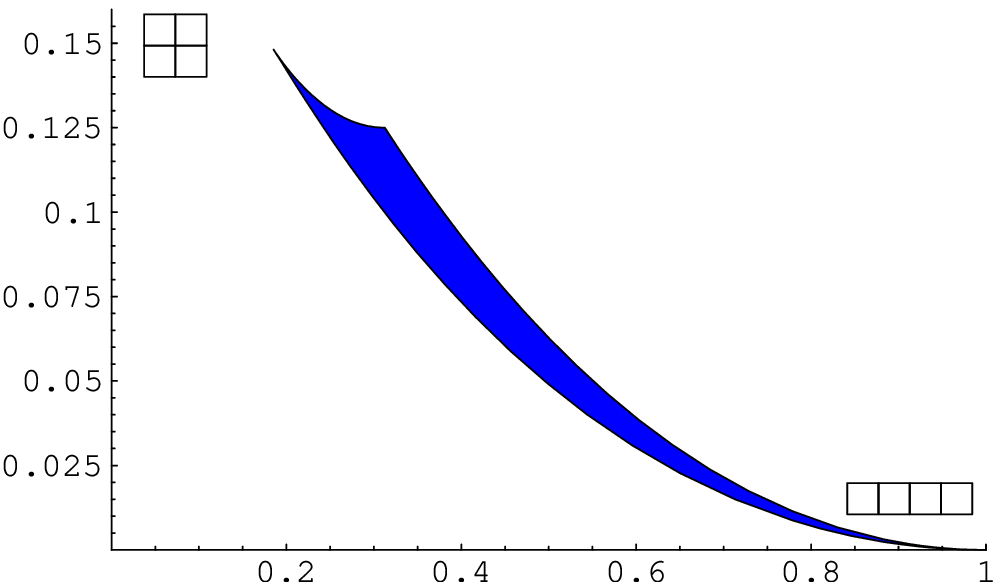}}
\caption{Left: the image $\tmap^k(\Delta(d))$ for $d=3$, $k=3$,
projected onto the coordinates $\rho_{(1^3)}$ and $\rho_{(2,1)}$. The
convex hull of $\tmap^k(\Delta(d))$ is a polytope (see
Example~\ref{example:wernerstatesvector}). Right: the image
$\tmap^k(\Delta(d))$ for $d=3$, $k=4$ projected onto $\rho_{(4)}$ and
$\rho_{(2,2)}$. The convex hull of this figure is not a polytope;
since it is equal to the projection of the convex hull of
$\tmap^k(\Delta(d))$, the latter set,
$\powerstates^4\cap\wernerstates^4$, cannot be a polytope.
\label{polytope}}
\end{figure}

Thus for $d=3$, $\powerstates^3\cap\wernerstates^3$ is a polytope (see
Figure \ref{polytope}), as in the previous example.  However, if the
number of diagrams with a given value of $k$ and $d$ exceeds $d$, the
situation is different:
\begin{theorem}\label{thm:notpolytopetheorem}
Let $k,d$ be such that $|\Par{k}{d}|>d$. Then the set
$\mytwirl^k(\powerstates^k)$ is not a polytope.
\end{theorem}
\begin{proof}

Let $X$ denote the subspace spanned by $\tmap^k(\myrq{q})$ for $q=1,
\ldots, d$, where we identify $\wernerstates^k$ with a subset of
$\mathbb{R}^{|\Par{k}{d}|}$, as in
Example~\ref{example:wernerstatesvector}. Since $|\Par{k}{d}|>d$,
there is a non-zero vector $v$ in $\mathbb{R}^{|\Par{k}{d}|}$ that is
orthogonal to $X$ with respect to the Euclidean scalar product in
$\mathbb{R}^{|\Par{k}{d}|}$.  Suppose $\tmap^k(r)$ lies in $X$ for all
$r \in \Delta(d)$. Then $\tmap^k(r).v=0$, for all $r$, so from Lemma
\ref{lem:schur-coefficient} we have for all $r\in \Delta(d)$
\be \label{zerorelation}
\sum_{\lambda\in\Par{k}{d}} (v_\lambda \dim V_\lambda) s_\lambda(r)=0\ ,
\ee
where $v=\sum_\lambda v_\lambda \rho_\lambda$. Since the Schur
polynomials are homogeneous, equation~(\ref{zerorelation}) extends
from $\Delta(d)$ to all $r$ with non-negative components, and
therefore all derivatives of the polynomial on the l.h.s. of this
equation are zero at the origin. Since every coefficient of this
polynomial is proportional to one of these derivatives, it must be
identically zero. But the Schur functions $s_\lambda$ form a basis for
the space of homogeneous symmetric polynomials of degree $k$ in $d$
variables, and therefore no such relationship can hold.

Therefore $\mytwirl^k(\powerstates^k)$ includes a point outside $X$. If
$\mytwirl^k(\powerstates^k)$ is a polytope, it has a vertex $w$ not in
$X$. Since $\mytwirl^k(\powerstates^k)$ is the convex hull of
$\tmap^k(\Delta(d))$, $w$ has the form $w=\tmap^k(a)$. As $w$ not in $X$,
$a$ is not a vertex of $\Delta(d)$, which implies that there is a line
segment in $\Delta(d)$ passing through $a$.  Because $\tmap^k$ is
smooth, the image under $\tmap^k$ of the line segment $t\mapsto a+t\xi$
has a tangent vector at the vertex $w$.  If this tangent vector does
not vanish, then we have a contradiction, since then the curve must
contain points outside the polytope $\mytwirl^k(\powerstates^k)$ in any
neighbourhood of $w$, however small.

It remains to show that, for any point $a \in \Delta(d)$ that is not a
vertex, there is a vector $\mathsf{\xi}\in\mathbb{R}^d$ such that
\begin{enumerate}
\item
the line segment $t\mapsto a+t\xi$ lies
within $\Delta(d)$ for  sufficiently small absolute values of the real
parameter $t$, and
\item
the derivative of $\tmap^k$ in the
direction $\mathsf{\xi}$ at the point $a$ has non-vanishing tangent
vector, i.e.  $\frac{\partial{\tmap^k(a+t\xi)}}{\partial t}|_{t=0}\neq 0$.
\end{enumerate}
It is
enough to show that the component of this tangent vector in some
direction $\mathsf{\tau}\in\mathbb{R}^{|\Par{k}{d}|}$ is non-vanishing,
i.e. that
\be \label{grad}
\frac{\partial{(\tau.\tmap^k(a+t\xi))}}{\partial t}\Big|_{t=0}
=\xi. (\nabla_r (\tau.\tmap^k(r)))\Big|_{r=a}\neq 0\ .
\ee
We choose $\mathsf{\xi}$ as follows: Suppose $a$ lies in the convex
hull of the $h$ vertices $\myrq{q_1}, \ldots, \myrq{q_h}$ of
$\Delta(d)$, arranged in increasing size of the index $q_i$, with $2
\le h \le d$. Thus
\be \label{face}
a=\sum_{i=1}^h u_i \myrq{q_i}, \mbox{  with } 0 < u_i <1 \mbox{ for } 1 \le i \le h.
\ee
Define
\begin{align}\label{eq:xidefinition}
\mathsf{\xi}&={q_1q_2 \over q_2-q_1}\left(\myrq{q_1}-\myrq{q_2}\right) \ ,\\ \nonumber
&=(\underbrace{1, \ldots, 1}_{q_1}, \underbrace{\beta,
  \ldots, \beta}_{q_2-q_1},0, \ldots, 0)\in\mathbb{R}^d
\end{align}
where $\beta={-q_1 \over q_2-q_1}$.
Then
$a+t\mathsf{\xi}$ lies within the convex hull of $\myrq{q_1}, \ldots,
\myrq{q_h}$, and hence in $\Delta(d)$, for small enough values of $|t|$.

To define $\mathsf{\tau}$, we use the fact
the monomial symmetric functions $m_\lambda$, for
$\lambda\in\Par{k}{d}$, also form a basis of the homogeneous
symmetric polynomials of degree $k$ in $d$ variables. In particular,
\[
m_{(d)}(r)=\sum r_i^d=\sum_\lambda \kappa_{\lambda,(d)} s_\lambda(r)\ ,
\]
where the coefficients $\kappa_{\lambda \mu}$ constitute the
transition matrix, which is given by the inverse of the matrix
of Kostka numbers~\cite{Mac79}. We now take
\[
\mathsf{\tau}=\sum_\lambda  {\kappa_{\lambda,(d)} \over \dim
  V_\lambda} \rho_\lambda\ ,
\]
which implies that
\[
\mathsf{\tau}.\tmap^k(r)=\sum r_i^d\ .
\]
From (\ref{grad}) and~\eqref{eq:xidefinition} therefore
\beastar
\xi. (\nabla_r (\tau.\tmap^k(r)))|_{r=a}&=&\sum_i\mathsf{\xi}_i \dd{\left(\sum_j r_j^d\right)}{r_i}\Big|_{r=a}\\
&=& d\sum_{i=1}^{q_1}a_i^{d-1}-{dq_1 \over
  q_2-q_1}\sum_{i=q_1+1}^{q_2}a_i^{d-1}\\
&>& 0\ ,
\eeastar
the last inequality holding because equation (\ref{face}) implies
\[
a_1 = \cdots =a_{q_1} > a_{q_1+1}= \cdots =a_{q_2}\ .
\]
The tangent vector at $a$ in the direction $\mathsf{\xi}$ is therefore
non-vanishing, which completes the proof.
\end{proof}

Figure \ref{polytope} shows an example where $d=3$, $k=4$ and
$|\Par{k}{d}|=4>d$.

\begin{figure}
\centerline{\includegraphics[width=0.4\textwidth]{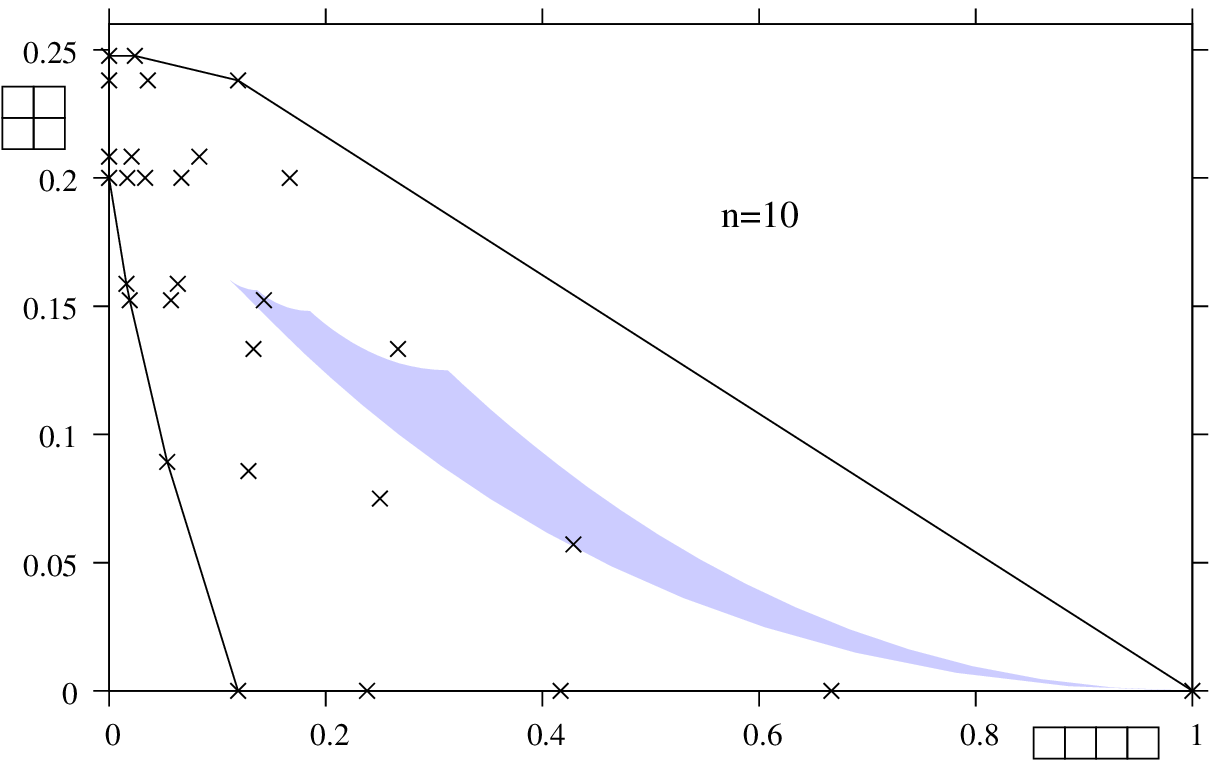}}
\vspace{0.1in}
\centerline{\includegraphics[width=0.4\textwidth]{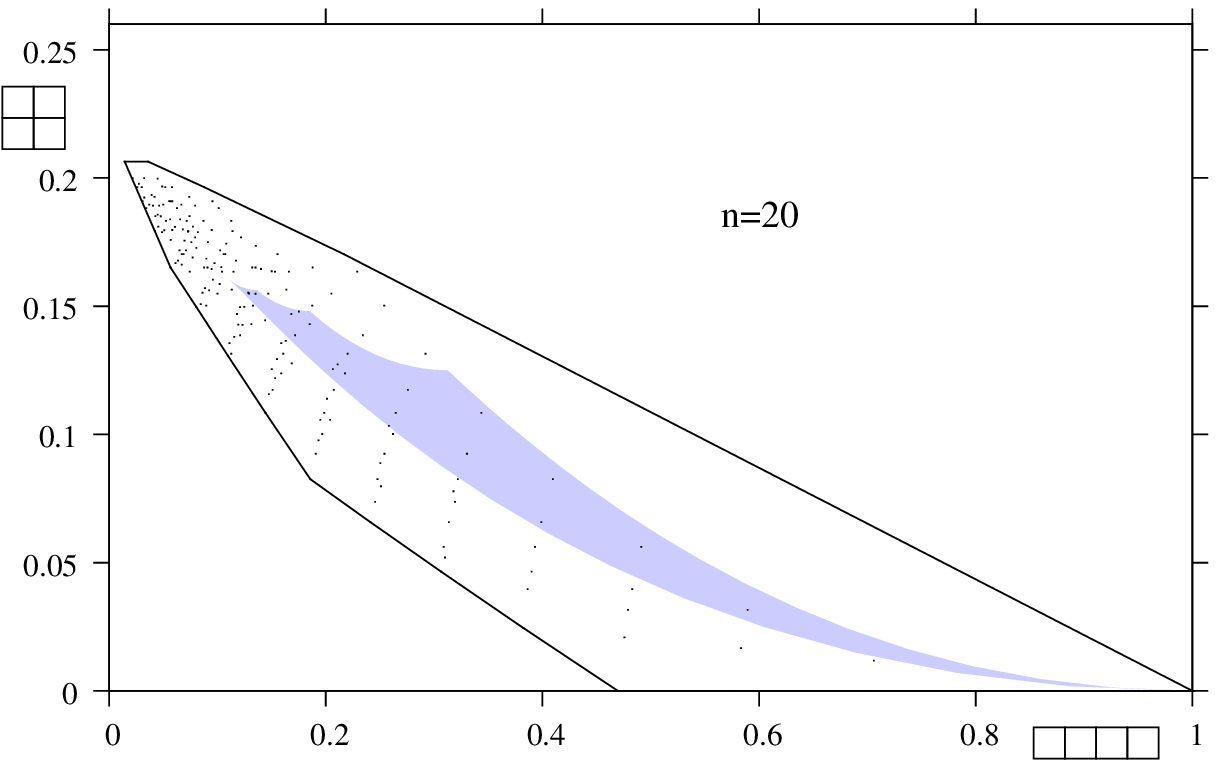}}
\vspace{0.1in}
\centerline{\includegraphics[width=0.4\textwidth]{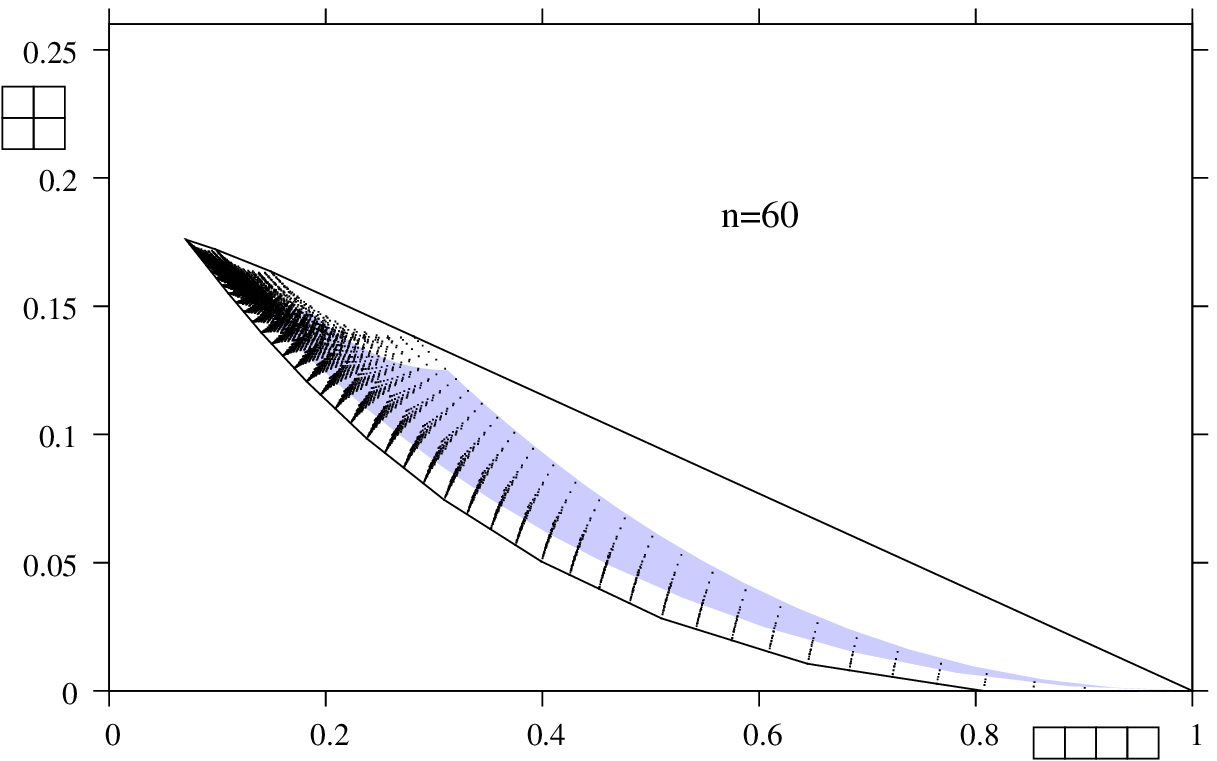}}
\caption{The figures show the image $\tmap^k(\Delta(d))$ (shaded
region) for $d=5$, $k=4$, projected onto the coordinates
$\rho_{(4)}$ and $\rho_{(2,2)}$. The image has a smooth convex
boundary, so $\powerstates^4\cap\wernerstates^4$ cannot be a
polytope. Also shown are the points obtained by tracing out $n-k$
systems from states in $\wernerstates^n$. Each point corresponds to
a diagram with $n=10$ boxes (top figure), $n=20$ (centre figure) and
$n=60$ boxes (bottom figure); the line segments demarcate the convex hull of all the
points. As expected, $\tmap^k(\Delta(d))$ is approximated more closely
as $n$ increases.} \label{striations}
\end{figure}

One might wonder whether Theorem \ref{thm:notpolytopetheorem} is
tight, in the sense that, for $|\Par{k}{d}| \le d$, the set
$\mytwirl^k(\powerstates^k)$ is a polytope. For $k=3$, $d=3$, where
$|\Par{k}{d}|=d$, we have seen that this is true. However, for $k=4$,
$d=5$, which also gives $|\Par{k}{d}|=d$, empirical evidence suggests
that $\mytwirl^k(\powerstates^k)$ is not a polytope, having a convex
boundary. This is shown in Figure \ref{striations}, which also plots
the images of traced-out states $\tr_{n-k} \rho^n_\lambda$ with $n=10$
and $n=60$ and shows how the approximation to $\mytwirl^k(\powerstates^k)$
improves as more systems are traced out; it also reveals some
intriguing striations in the case $n=60$, corresponding to diagrams
whose top rows are the same length.  Thus the characterisation of the
set $\powerstates^k\cap\wernerstates^k$ seems to be quite subtle, and
Werner states again uphold their reputation for exhibiting an
interesting variety of phenomena.

\section{Conclusions}\label{sec-conclusion}

Although the quantum de Finetti theorem is usually thought of as a
theorem about symmetric states, the unitary group shares the limelight
in the results described here. Our highest weight version of the de
Finetti theorem (Theorem \ref{thm:maindefinetti}) generalises the
usual symmetric-state version, but the extra generality almost comes
free; indeed, one could argue that the structure of the proof is made
clearer by taking the broader viewpoint. One can regard a highest
weight vector as the state in a representation that is as unentangled
as possible; this point of view has been taken by Klyachko
\cite{Klyachko02}. It is therefore natural to regard highest weight
vectors as analogues of product states, which is the role they have in
our theorem.

In the special case of symmetric states, our
Theorem~\ref{thm:definettisymmetric} gives bounds for the distance
between the $n$-exchangeable state $\rho^k$ and the set
$\powerstates^k$ of convex combinations of products $\sigma^{\otimes
k}$; these bounds are optimal in their dependence on $n$ and $k$,
the theorem giving an upper bound of order $k/n$ and there being
examples of states that achieve this bound (see
Theorem~\ref{diaconis-freedman}). The dependence of the bound on the
dimension $d$ is less clear, the theorem giving a factor of $d^2$
whereas in the classical case Diaconis and Freedman~\cite{DiaFre80}
obtained a bound with a dimension factor of order $d$.

Diaconis and Freedman also obtained a bound, $\frac{k(k-1)}{2n}$,
that is independent of the dimension. No such bound can exist for
quantum states, as Example \ref{alternating-state} shows; one can
find a state $\rho^n$ with the property that $\rho^2$, obtained by
tracing out all but two of the systems, lies at a distance at least $\textfrac{1}{2}$ from $\powerstates^2$. This example is a Werner state, in fact the
fully antisymmetric state on $d=n$ systems, and it is an
illustration of the usefulness of this family of states in giving
information about $\powerstates^k$.

Lemma~\ref{lem:wernerpartialtrace} shows that the shifted Schur
functions~\cite{OkoOls96} are closely connected with partial traces
of Werner states. The meaning of this connection needs to be further
explored: does the algebra of shifted symmetric functions have a
quantum-informational significance?

Another intriguing connection is with the theorem of Keyl and Werner
\cite{KeyWer01PRA}. They show that the spectrum of a state $\rho$
can be measured by carrying out a von Neumann measurement of
$\rho^{\otimes n}$ on the subspaces $U_\lambda \otimes V_\lambda$
in the Schur-Weyl decomposition of $(\mathbb{C}^d)^{\otimes n}$
(equation~\eqref{SchurWeyl}); if $\lambda$ is obtained, then $\bar \lambda=(\frac{\lambda_1}{n},\ldots,\frac{\lambda_d}{n})$ approximates the spectrum of $\rho$. Our theorem tells us
that $\rho^k=\tr_{n-k}\rho^n_\lambda$ can be
approximated by the twirled product $\sigma^{\otimes k}$, where
$\sigma$ has spectrum $\bar\lambda$. By the Keyl-Werner theorem, the
state $\tr_{n-k}\rho^n_\lambda$ must therefore project predominantly into subspaces
$U_\mu \otimes V_\mu$ with $\mu$ close to $\lambda$ in shape (but
rescaled by $k/n$). In this sense, tracing out a Werner state
approximately `preserves the shape' of its diagram. We can get an
intuition for why this should be by iterating the special case of
Lemma~\ref{lem:partialtraceproj} where one box is removed
(cf.~\cite[Proposition 4]{Audenaert04}). This shows that tracing
out is approximately equivalent, for large~$n$, to a process that
selects a row of a diagram with probability proportional to the
length of that row and then removes a box from the end of the row.

There have been many applications of the de Finetti theorem to topics
including foundational issues~\cite{fuchsschacksecond,Hudson81},
mathematical physics~\cite{fannes,raggiowerner} and quantum
information
theory~\cite{Ren05,bruncavesschack,dohertyetal,audenart,terhaldoherty,Ioannou06};
there have also been various
generalisations~\cite{DiaFre80,HudMoo76,stormer,fannesetal,raggiowerner,petz,cavesfuchsschack,fuchsschacksecond,KoeRen05,Ren05}.
We have taken one-and-a-half footsteps along this route.

\begin{acknowledgments} \vspace{-0.2cm}
We thank Aram Harrow and Andreas Winter for helpful discussions, and Ignacio Cirac and Frank Verstraete for raising
the question of how to approximate $n$-exchangeable states by
$m$-exchangeable states (see end of
section~\ref{sec-defin-werner}). We also thank the anonymous reviewers
for their helpful comments.

This work was supported by the EU project RESQ (IST-2001-37559) and
the European Commission through the FP6-FET Integrated Project
SCALA, CT-015714.  MC acknowledges the support of an EPSRC
Postdoctoral Fellowship and a Nevile Research Fellowship, which he
holds at Magdalene College Cambridge. GM acknowledges support from
the project PROSECCO~(IST-2001-39227) of the IST-FET programme of
the EC. RR was supported by Hewlett Packard Labs, Bristol.
\end{acknowledgments}

\end{document}